\documentclass[11pt]{article}
\usepackage{amssymb,amsmath,epsfig}
\usepackage{color}
\usepackage{graphicx}
\usepackage{mst-stylefile}
\usepackage{harvard}
\usepackage[makeroom]{cancel}
\usepackage{multirow,array}
\usepackage{caption}
\usepackage{epstopdf}
\usepackage{dsfont}
\usepackage{mathalfa}
\usepackage{fancyvrb}
\usepackage[utf8]{inputenc}
\usepackage{tikz}
\usetikzlibrary{shapes,arrows}
\usepackage{verbatim}
\usepackage{listings}% http://ctan.org/pkg/listings
\newcommand{\bbZ}{{\mathbb Z}}
\newcommand{\bbR}{{\mathbb R}}
\newcommand{\bbN}{{\mathbb N}}

\newcommand{\bbP}{{\mathbb P}}

\newcommand{\bbE}{{\mathbb E}}

\newcommand{\x}{\textbf{x}}

\newcommand{\supp}{\textnormal{supp}}

\newcommand{\Naturals}{\mathbb{N}}

\newcommand{\Reals}{\mathbb{R}}
\newcommand{\indicator}{\mathds{1}}

\renewcommand{\cite}{\citeyear}

\makeatletter
\def\moverlay{\mathpalette\mov@rlay}
\def\mov@rlay#1#2{\leavevmode\vtop{%
   \baselineskip\z@skip \lineskiplimit-\maxdimen
   \ialign{\hfil$\m@th#1##$\hfil\cr#2\crcr}}}
\newcommand{\charfusion}[3][\mathord]{
    #1{\ifx#1\mathop\vphantom{#2}\fi
        \mathpalette\mov@rlay{#2\cr#3}
      }
    \ifx#1\mathop\expandafter\displaylimits\fi}
\makeatother
\newcommand{\cupdot}{\charfusion[\mathbin]{\cup}{\cdot}}
\newcommand{\bigcupdot}{\charfusion[\mathop]{\bigcup}{\cdot}}
\begin{document}

\title{A spectral clustering-type algorithm for the consistent estimation of the Hurst distribution in moderately high dimensions
\thanks{H.W.~was partially supported by ANR-18-CE45-0007 MUTATION, France. G.D.'s long term visits to ENS de Lyon were supported by the school and by the Simons Foundation collaboration grant $\#714014$. The authors are very grateful to Viktor Todorov for bringing the database FRED-MD to their attention.}
}

\author{Patrice Abry \\ CNRS, ENS de Lyon, Laboratoire de Physique
\and   Gustavo Didier and Oliver Orejola\\ Mathematics Department\\ Tulane University
\and   Herwig Wendt\\ IRIT-ENSEEIHT, CNRS (UMR 5505),\\ Universit\'{e} de Toulouse, France}

\date{\today}

\bibliographystyle{agsm}

\maketitle

%-----------------------------------------------------------------------------------------------------------------

\begin{abstract}
Scale invariance (fractality) is a prominent feature of the large-scale behavior of many stochastic systems. In this work, we construct an algorithm for the statistical identification of the Hurst distribution (in particular, the scaling exponents) undergirding a high-dimensional fractal system. The algorithm is based on wavelet random matrices, modified spectral clustering and a model selection step for picking the value of the clustering precision hyperparameter. In a moderately high-dimensional regime where the dimension, the sample size and the scale go to infinity, we show that the algorithm consistently estimates the Hurst distribution. Monte Carlo simulations show that the proposed methodology is efficient for realistic sample sizes and outperforms another popular clustering method based on mixed-Gaussian modeling. We apply the algorithm in the analysis of real-world macroeconomic time series to unveil evidence for cointegration.
\end{abstract}

\section{Introduction}

\textit{Spectral clustering} is an embedding technique from the machine learning literature that has found numerous applications in data science (e.g., Hastie et al.\ \cite{hastie:tibshirani:friedman:2009}, Spielman and Teng \cite{spielman:teng:1996}, Ortega et al.\ \cite{ortega:frossard:kovacevic:moura:vandergheynst:2018}). A \textit{wavelet} is a unit $L^2(\bbR)$-norm function that annihilates polynomials (see \eqref{e:N_psi}). For a fixed (octave) $j \in \bbN \cup \{0\}$, a \textit{wavelet random matrix} is a multiscale sample covariance-like random matrix given by
\begin{equation}\label{e:W(a2^j)_intro}
\bbR^{p^2} \ni   {\boldsymbol{\mathcal W}}_n \equiv {\boldsymbol{\mathcal W}}(a(n)2^j) = \frac{1}{n_{a,j}}\sum^{n_{a,j}}_{k=1}{\mathcal D}(a(n)2^j,k)\hspace{0.5mm} {\mathcal D}(a(n)2^j,k)^{\top}.
\end{equation}
In \eqref{e:W(a2^j)_intro}, $^{\top}$ denotes transposition and $n_{a,j} = n/a(n)2^j$ is the number of wavelet-domain observations for a sample size $n$.  Each random vector ${\mathcal D}(a(n)2^j,k) \in \bbR^p$ is the \textit{wavelet transform} of a $p$-variate stochastic process $Y$ at dyadic scale $a(n)2^j$ and shift $k \in \bbZ$ (see \eqref{e:disc2}). A \textit{fractal} is an object or phenomenon that displays the property of \textit{self-similarity} across a range of scales, as characterized by the so-called \textit{scaling exponents} (Mandelbrot \cite{mandelbrot:1982}). Due to its intrinsic multiscale character and fine-tuned mathematical properties, the wavelet transform has been widely used in the study of fractals (e.g., Wornell \cite{wornell:1996}, Doukhan et al.\ \cite{doukhan:2003}, Massopust \cite{massopust:2014}). 

In this paper, we construct an algorithm for the estimation of the \textit{Hurst distribution} $\pi(dH)$ associated with a high-dimensional fractal system.  This involves estimating the Hurst modes (scaling exponents) and their probabilities, i.e., 
\begin{equation}\label{e:Hurst_modes_intro}
0 < \breve{H}_1 < \hdots < \breve{H}_r < 1  \quad \textnormal{and} \quad  \pi_i = \pi(\breve{H}_i), \quad  i = 1,\hdots,r,
\end{equation}
as well as $r \in \bbN$ itself. The estimation algorithm is based on \textit{wavelet random matrices}. It involves a modified \textit{spectral clustering} procedure for the estimation of $\pi(dH)$ given a precision hyperparameter $\varepsilon > 0$, and a \textit{model selection}-type procedure for picking $\varepsilon > 0$.  Following Abry et al.\ \cite{abry:didier:orejola:wendt:2024}, we assume measurements of the form
\begin{equation}\label{e:Y(ell)=P(n)X(ell)_intro}
\Reals^{p} \ni Y(t) = {\mathbf P}(n) X(t), \quad t \in \bbZ.
\end{equation}
In \eqref{e:Y(ell)=P(n)X(ell)_intro}, the $p \times p$ coordinates matrix ${\mathbf P}(n)$ is random, nonsingular and independent of $X(t)$. Moreover, each row of $X$ is, conditionally on ${\mathcal H} = H \in (0,1)$, an independent $H$-fractional Brownian motion, where each Hurst exponent ${\mathcal H}$ is picked independently from $\pi(dH)$ (see Definitions \ref{def:fBm} and \ref{def:rH-fBm}). We consider a \textit{moderately high-dimensional} framework where the \textit{sample size} $n$, the \textit{dimension} $p(n)$ and the \textit{scale} $a(n)2^j$ go to infinity while satisfying
\begin{equation}\label{e:three-fold_lim}
\frac{p(n) \cdot a(n)}{n} =  o\Big(\sqrt{\frac{a(n)}{n}}\Big), \quad \lim_{n \rightarrow \infty} \frac{a(n)}{n} = 0.
\end{equation}
In our main mathematical result (Theorem \ref{t:WRMSM_consistency}), we show that the proposed algorithm consistently estimates $\pi(dH)$. Simulation studies show that the algorithm is efficient over finite samples by comparison to a popular mixed-Gaussian clustering algorithm. We further apply the proposed algorithm in the analysis of real-world macroeconomic time series to unveil evidence for cointegration.\vspace{3mm}

In this paper, we combine two mathematical frameworks that are rarely considered jointly: $(i)$ cluster analysis and random matrix theory; $(ii)$ fractal analysis.

\textit{Cluster analysis} is a subject of broad interest in scientific and technological research today. Accordingly, it has been treated in the machine learning, statistics and computer science literatures, as well as in many engineering and societal applications (e.g., Meil\v{a} \cite{meila:2003}, Hastie et al.\ \cite{hastie:tibshirani:friedman:2009}, Di Marco and Navigli \cite{dimarco:navigli:2013}). The goal of any clustering procedure is to separate a set of data points in several groups based on some notion of \textit{similarity} (\textbf{n.b.}: not to be confused with ``self-similarity''). In particular, \textit{spectral clustering} is a technique that reinterprets the data in the form of a graph (e.g., Kannan et al.\ \cite{kannan:vempala:vetta:2004}, von Luxburg \cite{vonLuxburg:2007}, Filippone et al.\ \cite{filippone:camastra:masulli:rovetta:2008}, Rohe et al.\ \cite{Rohe:Chatterjee:Yu:2011}, Vershynin \cite{vershynin:2018}).  This graph is subsumed under the so-named \textit{graph Laplacian (matrix)} (e.g., Chung \cite{chung:1997}; see Definition \ref{d:adjacency_mat}). Ultimately, the graph Laplacian bridges the gap between useful low-dimensional representations, such as graphs, and complex objects, such as manifolds (He and Yau \cite{he:yau:2022}). These representations are particularly powerful when dealing with non-convex domains (e.g., see Shi and Malik \cite{Shi:Malik:2000} on computer vision). The robustness of spectral clustering with respect to perturbations is studied in Ng et al.\ \cite{ng:jordan:weiss:2001}. Its consistency is further established under broad conditions in von Luxburg et al.\ \cite{vonLuxburg:Belkin:Bousquet:2008} assuming independent sampling.

Now recall that the emergence of a \textit{fractal} is typically the signature of a physical mechanism that generates \textit{scale invariance} (e.g., West et al.\ \cite{west:brown:enquist:1999}, Zheng et al.\ \cite{zheng:shen:wang:li:dunphy:hasan:brinker:su:2017}, He \cite{he:2018}, Shen et al.\ \cite{shen:stoev:hsing:2022}). On one hand, scale invariance is empirically detected in a wide range of natural and social phenomena such as in turbulence (Kolmogorov~\cite{Kolmogorovturbulence}, Parisi and Frisch \cite{parisi:frisch1985}), climate studies (Isotta et al.\ \cite{isotta:etal:2014}), dendrochronology (Bai and Taqqu~\cite{bai:taqqu:2018}), neuroscience (Beggs \cite{beggs:2008}) and hydrology (Benson et al.\ \cite{benson:baeumer:scheffler:2006}). On the other hand, the conceptual characterization of scale invariance is a major theme in modern research. This is so in many analytical frameworks, including those of high-dimensional learning (Holl et al.\ \cite{holl:koltun:thuerey:2022}, Hu and Lu \cite{hu:lu:2022}), renormalization (Cardy \cite{cardy:1996}, O'Malley and Cushman \cite{omalley:cushman:2012}, Bauerschmidt et al.\ \cite{bauerschmidt:brydges:slade:2019}), criticality and phase transitions (Christensen and Moloney \cite{christensen:moloney:2005}, Sornette \cite{sornette:2006}) and percolation (Smirnov \cite{smirnov:2001}, Duminil-Copin and Tassion \cite{duminil-copin:tassion:2016}). In particular, scale invariance is a recurrent theme in the fundamental pursuit of \textit{universality} (Soshnikov \cite{soshnikov:1999}, Tao \cite{tao:2012}, Seddik et al.\ \cite{seddik:louart:couillet:tamaazousti:2021}, Lu and Yau \cite{lu:yau:2023}).

In the absence of a classical \textit{characteristic scale} (e.g., Reif \cite{reif:2009}), the modeling of scale-invariant (fractal) systems revolves around \textit{scaling exponents}, which determine the behavior of the system across a continuum of scales. In ``Big Data" applications (Briody \cite{briody:2011}), fractality usually involves \textit{multiple} scaling laws evolving along \textit{non-canonical} coordinate axes. For example, in econometrics, the detection of distinct scaling laws in multivariate fractional time series is indicative of the key property of \textit{cointegration} -- namely, the existence of meaningful and statistically useful long-run relationships among the individual series (e.g., Engle and Granger \cite{engle:granger:1987}, NobelPrize.org \cite{nobelprize:2003}, Hualde and Robinson \cite{hualde:robinson:2010}, Shimotsu \cite{shimotsu:2012}). In Abry and Didier \cite{abry:didier:2018:n-variate} and Lucas et al.\ \cite{lucas:abry:wendt:didier:2023}, multiple scaling exponents are uncovered in the contexts of network traffic and neuroscientific data modeling, respectively. Other areas involving multiple scaling laws include neuroscience and fMRI imaging (Li et al.\ \cite{li:pluta:shahbaba:fortin:ombao:baldi:2019}, Gotts et al.\ \cite{gotts:gilmore:martin:2020}, Beggs \cite{beggs:2022}), climate science (Schmith et al.\ \cite{schmith:johansen:thejll:2012}), high-dimensional time series (Merlev\`{e}de and Peligrad \cite{merlevede:peligrad:2016}, Chan et al.\ \cite{chan:lu:yau:2017}, Alshammri and Pan \cite{alshammri:pan:2021}) as well as signal processing (Comon and Jutten \cite{comon:jutten:2010},  Abry et al.\ \cite{abry:didier:li:2019}).

%From a theoretical standpoint, a $\bbR^r$-valued stochastic process $X$ is called \textit{operator self-similar} (o.s.s.; Laha and Rohatgi \cite{laha:rohatgi:1981}, Hudson and Mason \cite{hudson:mason:1982}) if it exhibits the scaling property
%\begin{equation}\label{e:def_ss}
%\{X(ct)\}_{t\in\bbR} \stackrel {\textnormal{f.d.d.}}{=} \{c^{\mathbf H} X(t)\}_{t\in\bbR}, \quad c>0.
%\end{equation}
%In \eqref{e:def_ss}, ${\mathbf H}$ is some (Hurst) matrix whose eigenvalues have real parts lying in the interval $(0,1)$ and $c^{\mathbf H} := \exp\{\log(c) {\mathbf H}\} = \sum^{\infty}_{k=0} \frac{(\log(c) {\mathbf H})^k}{k!}$. \GD{Relation \eqref{e:def_ss} implies that $X$ displays \textit{multiple scaling laws} driven by the eigenvalues of ${\mathbf H}$.} A canonical model for multivariate fractal systems is \textit{operator fractional Brownian motion} (ofBm), namely, a Gaussian, o.s.s., stationary-increment stochastic process . OfBm is the natural multivariate generalization of the classical fractional Brownian motion.

From a different perspective, \textit{random matrices} have emerged as one essential analytical framework in mathematical physics (e.g., Mehta and Gaudin \cite{mehta:gaudin:1960}, Dyson \cite{dyson:1962}, Ben Arous and Guionnet \cite{benarous:guionnet:1997}, Soshnikov \cite{soshnikov:1999}, Anderson et al.\ \cite{anderson:guionnet:zeitouni:2010}, Erd\H{o}s et al.\ \cite{erdos:yau:yin:2012}) as well as in high-dimensional statistics and machine learning (e.g., Bai and Silverstein \cite{bai:silverstein:2010}, Tao and Vu \cite{tao:vu:2012}, Boucheron et al.\ \cite{boucheron:lugosi:massart:2013}, Xia et al.\ \cite{xia:qin:bai:2013}, Paul and Aue \cite{paul:aue:2014}, Giraud \cite{giraud:2015}, Yao et al.\ \cite{yao:zheng:bai:2015}, Wainwright \cite{wainwright:2019}, Couillet and Liao \cite{couillet:liao:2022}). In particular, \textit{wavelet random matrices} (WRMs) provide a natural framework for the study of high-dimensional stochastic dynamics over large scales (i.e., over the long term). This is so because the scaling laws hidden in measurements eventually emerge among the eigenvalues of WRMs. To the very best of our knowledge, WRMs were first studied by the authors and collaborators in a series of papers (e.g., Abry and Didier \cite{abry:didier:2018:dim2}, Orejola et al.\ \cite{orejola:didier:wendt:abry:2022}, Abry, Boniece et al.\ \cite{abry:boniece:didier:wendt:2023:regression,abry:boniece:didier:wendt:2024}). 

The first results on eigenvalues in the \textit{bulk} of the spectrum of large WRMs were provided in Abry, Didier et al.\ \cite{abry:didier:orejola:wendt:2024}. In the context of that paper, one assumes measurements of the form \eqref{e:Y(ell)=P(n)X(ell)_intro}. Then, it can be shown that 
\begin{equation}\label{e:Y_scales}
\{Y(ct)\}_{t \in \bbR} \stackrel{\textnormal{f.d.d.}}= \{ {\mathbf P}(n)\hspace{0.5mm} c^{\mathds{H}_n} \hspace{0.5mm}{\mathbf P}(n)^{-1} \hspace{0.5mm}Y(t)\}_{t \in \bbR}, \quad \textnormal{for any $c > 0$} 
\end{equation}
(cf.\ Abry, Didier et al.\ \cite{abry:didier:orejola:wendt:2024}, Lemma C.12). In \eqref{e:Y_scales}, $c^{\mathds{H}_n}$ is a $p \times p$ diagonal matrix containing copies of $c^{\mathcal H} \in (0,\infty)$ along the main diagonal, where ${\mathcal H} \sim \pi(dH)$ (cf.\ \eqref{e:mathds_hn}). Hence, the process \eqref{e:Y(ell)=P(n)X(ell)_intro} is scale-invariant and displays \textit{multiple} scaling laws along \textit{non-canonical} axes. For this reason, it provides a natural model for high-dimensional fractal systems.

\begin{figure}
    \centering
    \includegraphics[scale=0.25]{"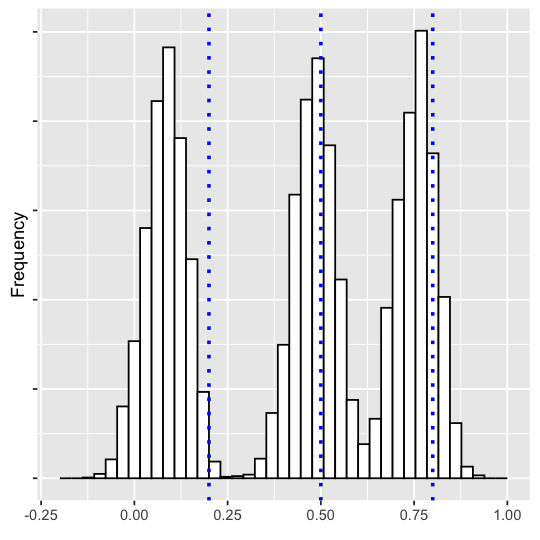"}
    \includegraphics[scale=0.25]{"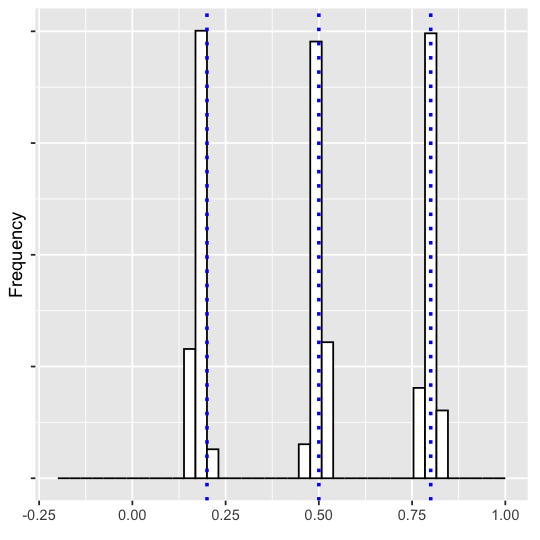"}
 \caption{ \textbf{The distribution of the (rescaled logarithmic) wavelet e.s.d.\ in the three-way limit \eqref{e:three-fold_lim} (see Abry, Didier et al.\ \protect\cite{abry:didier:orejola:wendt:2024}).} A Monte Carlo study displays a tri-modal distribution emerging in the rescaled logarithmic wavelet e.s.d.\ in the three-way limit \eqref{e:three-fold_lim} (\textbf{n.b.}: after applying an affine transformation, the results are shown on the same scale as that of the distribution $\pi(dH)$). In the depicted simulation study based on $1000$ realizations, $\pi(dH)$ is a discrete uniform distribution with support $\{0.2,0.5,0.8\}$. For the left and right plots, respectively, $ (\textnormal{sample size}, \textnormal{scale},\textnormal{dimension} )  = (2^{10}, 2^4, 2^3)$ and $(2^{18}, 2^6, 2^6)$.  Wavelet log-eigenvalues weighted over multiple scales were used for enhanced (``debiased") finite-sample convergence (cf.\ Abry and Didier~\protect\cite{abry:didier:2018:n-variate} and Abry et al.\ \protect\cite{abry:boniece:didier:wendt:2023:regression}).  }
    \label{fig:histogram}
\end{figure}

In Abry, Didier et al.\ \cite{abry:didier:orejola:wendt:2024}, it is then established that the logarithmic empirical spectral distribution (log-e.s.d.) of ${\boldsymbol {\mathcal W}}_n$ converges weakly, in probability, to the distribution of the random variable $2\mathcal{H}+1$ (see Figure \ref{fig:histogram}; see also Theorem \ref{t:main_theorem_discrete} in this paper for the precise statement). When taking limits, one considers the moderately high-dimensional regime \eqref{e:three-fold_lim}, where $n$ grows faster than $p(n)\cdot a(n)$  (see \eqref{e:p(n),a(n)_conditions}; cf.\ Wang et al.\ \cite{wang:aue:paul:2017}). In particular, taking the three-way limit \eqref{e:three-fold_lim} allows for probing the \textit{large-scale} properties (i.e., long-term behavior) of the model \eqref{e:Y(ell)=P(n)X(ell)_intro} \textit{in (moderately) high dimensions}.

\begin{figure}
    \centering
    \includegraphics[scale=0.35]{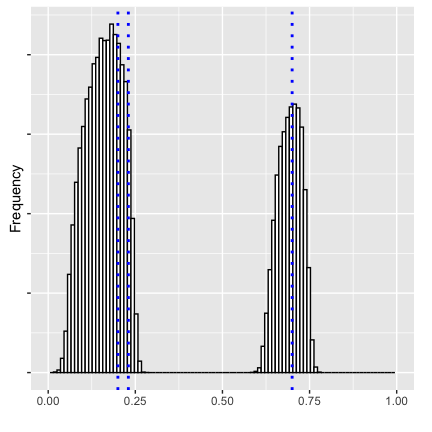}
 \caption{ \textbf{How many Hurst modes?} Notwithstanding the good asymptotic properties of the wavelet log-e.s.d.\ (see Theorem \ref{t:main_theorem_discrete}), over finite samples the information provided may be ambiguous for differences between Hurst modes that are small relative to the sample size. For the sake of illustration, the log-eigenvalue plot displayed might suggest the existence of two modes $0 < \breve{H}_1 < \breve{H}_2 < 1$ appearing with probabilities $\pi(\breve{H}_1) > \pi(\breve{H}_2)$. In truth, though, it was simulated based on a discrete uniform distribution on $\{0.25,0.29,0.7\}$, as indicated by the vertical dotted lines.}
    \label{fig:3HurstClose}
\end{figure}

Even though the log-e.s.d.\ of ${\boldsymbol {\mathcal W}}_n$ ultimately converges to the distribution of $2 {\mathcal H}+1$, for modeling purposes such convergence can be a delicate issue both in statistical and numerical senses. Namely, given a histogram based on a finite sample of wavelet log-eigenvalues, how can modelers arrive at accurate conclusions on the true number and distribution of Hurst modes? (see Figure \ref{fig:3HurstClose} for an illustration) This challenge got first tackled in Orejola et al.\ \cite{orejola:didier:wendt:abry:2022} by means of a hypothesis test for unimodality (i.e., for whether or not $r = 1$ in \eqref{e:Hurst_modes_intro}).

In this paper, we combine wavelet random matrices, spectral clustering and model selection to put forth an algorithm for estimating the scaling structure -- i.e., the Hurst distribution $\pi(dH)$ -- of the high-dimensional fractal system \eqref{e:Y(ell)=P(n)X(ell)_intro}. The algorithm -- hereinafter referred to as \textbf{WRMSM} (\textbf{W}avelet \textbf{R}andom \textbf{M}atrix, modified \textbf{S}pectral clustering and \textbf{M}odel selection) -- converts the estimation of the asymptotic wavelet log-e.s.d.\ into the quantifiable detection and estimation of the Hurst modes and their probabilities over finite sample sizes. It comprises two main components. Given a precision hyperparameter value $\varepsilon > 0$, a \textit{Hurst distribution $\varepsilon$-precision estimation subroutine} (\textbf{HD$\varepsilon$ES}) clusters the wavelet log-e.s.d.\ by means of a modified spectral clustering-type procedure and constructs estimators of the Hurst distribution. Then, \textbf{WRMSM} uses the so-called \textit{intra-cluster standard deviation} to conduct model selection as a function of the hyperparameter $\varepsilon$ utilized by \textbf{HD$\varepsilon$ES}. Notwithstanding the \textit{dependence} among wavelet log-eigenvalues, the resulting algorithm is provably consistent, as established in our main mathematical result (Theorem \ref{t:WRMSM_consistency}). Interestingly, \textbf{HD$\varepsilon$ES} per se provides consistent estimation of the Hurst distribution $\pi(dH)$ (Proposition \ref{t:fixed_epsilon_consistency}). Nevertheless, the model selection carried out by  \textbf{WRMSM} addresses the problem of hyperparameter choice and improves the overall finite-sample estimation performance (on rates of convergence, see Remark \ref{r:rates_of_convergence}, ($ii$)). Thorough computational studies demonstrate that the algorithm displays quantifiable finite-sample robustness and superior performance by comparison to the classical clustering method called \textit{Gaussiam mixture model} (\textbf{GMM}; see Section \ref{s:computational_studies}). As an application of the mathematical and computational results, we investigate the presence of the property of cointegration in a database of monthly frequency macroeconomic time series held by the St.\ Louis Federal Reserve Bank (see Section \ref{s:Application}).

The paper is organized as follows. In Section \ref{s:preliminaries}, we lay out the notation as well as basic concepts of wavelets, wavelet random matrices and graphs. In Section \ref{s:main_results}, we establish the main mathematical result of the paper, i.e., the consistency of \textbf{WRMSM}. In Section \ref{s:computational_studies}, we describe computational studies of the finite-sample estimation performance. In Section \ref{s:Application}, we conduct the data analysis. In Section \ref{s:conclusion}, we provide conclusions and a discussion of open problems. All proofs and auxiliary results can be found in the Appendix.

\section{Preliminaries and assumptions}\label{s:preliminaries}

Hereinafter, $|{\mathcal X}|$ denotes the cardinality of a set ${\mathcal X} \subseteq \bbR^d$, $d \in \bbN$. For $\x \in \bbR^d$ and $\varepsilon > 0$, ${\mathcal B}_{\varepsilon}(\x)$ denotes the open ball around $\x$ with radius $\varepsilon$. All with respect to the field $\bbR$, ${\mathcal M}(m_1,m_2,\bbR)$ and ${\mathcal M}(n)={\mathcal M}(n,\bbR)$ are the vector spaces, respectively, of all $m_1 \times m_2$ matrices and $n \times n$ matrices, $GL(n,\bbR)$ is the general linear group (invertible matrices), and $O(n)$ is the orthogonal group (i.e., matrices $O \in {\mathcal M}(n)$ such that $OO^{\top} = I$). Also, ${\mathcal S}(n,\bbR)$ and ${\mathcal S}_{\geq 0}(n,\bbR)$ denote, respectively, the space and the cone of $n \times n$ symmetric and symmetric positive semidefinite matrices, whereas $I = I_{n}$ denotes the $n\times n$ identity matrix. Given a collection of scalars $x_1,\hdots,x_n$,
\begin{equation}\label{e:x(1)=<...=<x(n)}
x_{(1)} \leq \hdots \leq x_{(n)}
\end{equation}
denotes the associated ordered $n$-tuple.

\subsection{Wavelet random matrices}

Recall that a \textit{wavelet} $\psi$ is a unit $L^2(\bbR)$-norm function that annihilates polynomials (see \eqref{e:N_psi}). Throughout the paper, we make use of a \textit{wavelet multiresolution analysis} (MRA; see Mallat \cite{mallat:1999}, chapter 7), which decomposes $L^2(\mathbb{R})$ into a sequence of \textit{approximation} (low-frequency) and \textit{detail} (high-frequency) subspaces $V_j$ and $W_j$, respectively, associated with different scales of analysis $2^{j}$, $j \in \bbZ$. In particular, given a wavelet $\psi$, there is a related \textit{scaling function} $\phi \in L^2(\bbR)$. Appropriate rescalings and shifts of $\phi$ and $\psi$ form bases for the subspaces $V_j$ and $W_j$, respectively (see Mallat \cite{mallat:1999}, Theorems 7.1 and 7.3).

In almost all mathematical statements, we make assumptions ($W1-W3$) on the underlying wavelet MRA. Such assumptions are standard in the wavelet literature and are accurately described in Section \ref{s:assumptions_on_the_MRA}. In particular, we make use of a compactly supported wavelet basis.

We now define the \textit{wavelet transform} of a vector-valued stochastic process in the discrete-time setting. So, consider the process
\begin{equation}\label{e:finite_sample}
\{Y(k)\}_{k=1,\hdots,n} \subseteq \bbR^p,
\end{equation}
associated with the starting scale $2^j = 1$ (or octave $j = 0$). In particular, $n \in \bbN$ in \eqref{e:finite_sample} denotes the sample size. Starting from \eqref{e:finite_sample}, we suppose the wavelet transform vector ${\mathcal D}(2^j,k)$ of the high-dimensional process $Y$ stems from Mallat's pyramidal algorithm (Mallat \cite{mallat:1999}, chapter 7). It is given by the convolution
\begin{equation}\label{e:disc2}
\bbR^p \ni {\mathcal D}(2^j,k) := \sum_{\ell\in\bbZ} Y(\ell)h_{j,2^jk-\ell}, \quad j \in \bbN \cup \{0\},
\end{equation}
where we use the convention $Y(k) = 0$ for $k \notin \{1,\hdots,n\}$ and the filter terms are defined by
\begin{equation}\label{e:hj,l}
\bbR \ni h_{j,\ell} =2^{-j/2}\int_\bbR\phi(t+\ell)\psi(2^{-j}t)dt, \quad \ell \in \bbZ.
\end{equation}
A more detailed description of Mallat's algorithm is provided in Section \ref{s:Mallats_algorithm}. The associated \textit{wavelet random matrix} is given by expression \eqref{e:W(a2^j)_intro}, namely,
\begin{equation}\label{e:W(a(n)2^j)_intro}
{\mathcal S}_{\geq 0}\big(p(n),\bbR \big) \ni {\boldsymbol {\mathcal W}}_n \equiv {\boldsymbol {\mathcal W}}\big(a(n)2^j\big) := \frac{1}{n_{a,j}} \sum^{n_{a,j}}_{k=1} {\mathcal D}\big(a(n)2^j,k\big)\hspace{0.5mm}{\mathcal D}\big(a(n)2^j,k\big)^{\top}, \quad n_{a,j} = \frac{n}{a(n)2^j}.
\end{equation}
In \eqref{e:W(a(n)2^j)_intro}, $n_{a,j}$ is the \textit{effective sample size}, i.e., the number of wavelet-domain observations available at scale $a(n)2^j$.

\subsection{Measurements and assumptions}

In regard to the underlying stochastic framework, first consider the following two definitions.

\begin{definition}\label{def:fBm}
Fix $0 < H < 1$. A standard \textit{fractional Brownian motion} (fBm) is the Gaussian stochastic process $\{B_H(t)\}_{t \i \bbR}$ whose covariance function is given by
\begin{equation}\label{e:fBm_cov}
\bbE B_H(s) B_H(t) = \frac{1}{2} \big\{ |s|^{H} + |t|^{H} - |t- s|^{H}\big\}, \quad s,t \in \bbR.
\end{equation}
\end{definition}
Relation \eqref{e:fBm_cov} and Gaussianity imply that fBm is \textit{self-similar}, i.e., it satisfies the property
$$
\{ B_H(ct) \}_{t \in \bbR} \stackrel{\textnormal{f.d.d.}}= \{ c^{H}B_H(t) \}_{t \in \bbR}, \quad c > 0
$$
(Embrechts and Maejima \cite{embrechts:maejima:2002},  Pipiras and Taqqu \cite{pipiras:taqqu:2017}).

\begin{definition}\label{def:rH-fBm}
Let $\pi(dH)$ be a probability measure such that $\pi(0,1) = 1$. The univariate stochastic process
\begin{equation}\label{e:X_h(t)_def}
X_\mathcal{H} = \{X_{\mathcal{H}}(t)\}_{t \in \bbR}
\end{equation}
is called a \textit{random Hurst (exponent)--fractional Brownian motion} (rH--fBm) when, conditionally on some value $\mathcal{H} = H$ picked from $\pi(d H)$, $X_{\mathcal{H}}$ is a standard fBm with Hurst exponent
\begin{equation}\label{e:H_in_(0,1]}
H \in (0,1).
\end{equation}
\end{definition}
The basic properties of rH-fBm can be found in Abry et al.\ \cite{abry:didier:orejola:wendt:2024}, Lemma C.10 (see also Balcerek et al.\ \cite{balcerek:burnecki:thapa:wylomanska:chechkin:2022}).\\

So, throughout this manuscript, we make use of the following assumptions on the measurements. For expository purposes, we first state the assumptions, and then provide some interpretation. In the assumptions, ${\mathcal T} = \{1,\hdots,n\}$.

\medskip

\noindent  {\sc Assumption $(A1)$}:  $\pi(d H)$ is a distribution such that
\begin{equation}\label{e:pi(dh)}
\supp \ \pi(d H) = \{\breve{H}_1,\hdots,\breve{H}_{r} \}, \quad \varpi :=\min \{\supp \ \pi (d H)\},
\end{equation}
where $r \in \Naturals$ and \eqref{e:Hurst_modes_intro} holds. In addition, for $i = 1,\hdots,r$ and $\pi_i$ as in \eqref{e:Hurst_modes_intro}, 
\begin{equation}\label{e:difference_of_probabilites_bounded}
\min_{i=1,\hdots,r} \pi_i  > \max_{i = 1,\hdots, r}\{ \pi_{i+1} - \pi_{i} \}.
\end{equation}

\medskip

\noindent  {\sc Assumption $(A2)$}: For each $n \in \Naturals$, $p=p(n)$ independent rH--fBm sample paths are generated based on (independently) sampling from $\pi(d H)$ as in  \eqref{e:pi(dh)}. In particular, when restricted to discrete time, a total of $p \times n$ entries is available. The associated $p$-variate (latent) stochastic process is denoted by $\{X(t)\}_{t \in {\mathcal T}}$.

\medskip

\noindent {\sc Assumption $(A3)$}: For $p = p(n)$ and $\{X(t)\}_{t \in {\mathcal T}}$ as in assumption $(A2)$, the measurements have the form
\begin{equation}\label{e:Y=PX}
\Reals^{p(n)} \ni Y(t) = {\mathbf P}(n) X(t), \quad t \in {\mathcal T}.
\end{equation}
In \eqref{e:Y=PX}, the so-named \textit{coordinates matrix} is a random matrix ${\mathbf P}(n) \in GL\big( p(n),\Reals \big)$ that is independent of $X(t)$.

\medskip

\noindent {\sc Assumption $(A4)$}: Fix $j \in \bbN \cup \{0\}$. The dimension $p(n)$ and the scaling factor $a(n)$ (cf.\ $n_{a,j}$ as in \eqref{e:W(a(n)2^j)_intro}) satisfy the relations
\begin{equation}\label{e:p(n),a(n)_conditions}
a(n) \leq \frac{n}{2^j}, \quad \frac{a(n)}{n} + \frac{n}{a(n)^{1+ 2\varpi}} \rightarrow 0, \quad p(n) < \frac{n}{a(n)2^j}, \quad \infty \leftarrow p(n) = o\Big(\sqrt{\frac{n}{a(n)}}\Big),
\end{equation}
as $n\to \infty$, where $\varpi$ is as in \eqref{e:pi(dh)}.

\medskip

\noindent {\sc Assumption $(A5)$}: For the random matrix ${\mathbf P}(n)$ as in \eqref{e:Y=PX},

\begin{equation}\label{e:constraints_on_coordinate_matrix}
    \frac{\log \sigma_1(\mathbf{P}(n))}{\log a(n)} \stackrel{\bbP}\to 0,\quad \frac{\log \sigma_p(\mathbf{P}(n))}{\log a(n)} \stackrel{\bbP}\to 0, \quad n \rightarrow \infty.
\end{equation}

\medskip

Assumption $(A1)$ defines the distribution of Hurst exponents. In particular, condition \eqref{e:difference_of_probabilites_bounded} is \textit{not} mathematically essential. It simply accounts for the fact that one cannot generally expect graph Laplacians computed from real data to display exactly null eigenvalues (cf.\  \textbf{HD$\varepsilon$ES}, Section \ref{s:main_results}). Assumption $(A2)$ postulates the latent $p(n)$-variate stochastic process $X$ as a collection of independent rH-fBms. Assumption $(A3)$ describes the observed process $Y$, where the \textit{unknown} random coordinates matrix ${\mathbf P}(n)$ determines the directions of the multiple (random) scaling relations stemming from the latent process $X$. In turn, assumption $(A4)$ controls the divergence rates among $n$, $a(n)$ and $p(n)$ in the three-way limit. In particular, it states that the three-component ratio $\frac{p(n)\cdot a(n)}{n}$ must converge to a constant $c = 0$ (see \eqref{e:three-fold_lim}). This establishes a moderately high-dimensional regime (cf.\ the traditional ratio $\lim_{n \rightarrow \infty} \frac{p(n)}{n}$ for sample covariance matrices).\\

Throughout this manuscript, for each $n \in \Naturals$ it will be useful to write 
\begin{equation}\label{e:mathds_hn}
\mathds{H}_n  = \textnormal{diag}(H_1,\hdots,H_{p(n)})\in {\mathcal M}\big(p(n)\big)
\end{equation}
to denote a diagonal random matrix whose (main diagonal) entries correspond to the sampled Hurst exponents as described in assumption $(A2)$. Also, whenever convenient we write
\begin{equation}\label{e:a=a(n)_or_p=p(n)}
a = a(n)\textnormal{ or }p = p(n).
\end{equation}

\begin{remark}
    Due to the many uses of the letter ``$H$'' throughout the paper, for the readers' convenience we provide Table \ref{tab:notation_for_h} to help them keep track of the notation.
\begin{table}[]
    \centering
        \begin{tabular}{cclc}
        \textbf{notation} & \textbf{domain} & \textbf{description} & \textbf{defined in} \\ \hline
 %       $\mathbf{H}$ & ${\mathcal M}(p)$ & Hurst matrix &  \eqref{e:def_ss}    \\
        $\mathds{H}_n$  & ${\mathcal M}(p)$ & diagonal matrix whose main diagonal  & \eqref{e:mathds_hn} \\
                     &                   & entries are picked from $\pi(dH)$ &                \\
        $\mathsf{H}_p$ & set & diagonal entries of $\mathds{H}_n$ & \eqref{e:sampled_exponenets} \\
        $\mathcal{H}$ & $(0,1)$  & random Hurst exponent & \eqref{e:X_h(t)_def} \\
       $H$ & $(0,1)$  & particular (deterministic) instance of $\mathcal{H}$  & \eqref{e:H_in_(0,1]} \\
       $\widehat{H}_\ell$ & $\mathbb{R}$  & rescaled and shifted wavelet log-eigenvalue & \eqref{e:scaling_estimate} \\
       $\widehat{\mathsf{H}}_p$ & set  & collection of $\widehat{H}_\ell$ & \eqref{e:scaling_estimate} \\
       $\breve{H}_i$ & $(0,1)$  & value in $\textnormal{supp}\hspace{0.5mm}\pi(dH)$  & \eqref{e:pi(dh)}\\
        $\widehat{\breve{H}}_i$ & $(0,1)$  & estimator of $\breve{H}_i$ & \eqref{e:H-breve-hat_pi-hat} \\\hline
    \end{tabular}
    \caption{Different uses of the letter ``$H$" throughout the paper.}
    \label{tab:notation_for_h}
    %This is silly \label must be used after caption.
\end{table}
\end{remark}

%\noindent {\sc Assumption ($A4$)}: Fix $j \in \bbN \cup \{0\}$. The dimension $p(n)$ and the scaling factor $a(n)$ (see \eqref{e:W(a2^j)_intro}) satisfy the relations
%\begin{equation}\label{e:p(n),a(n)_conditions}
%a(n) \leq \frac{n}{2^j}, \quad \frac{a(n)}{n} + \frac{n}{a(n)^{1+ 2\varpi}} \rightarrow 0, \quad p(n) < \frac{n}{a(n)2^j},
%\end{equation}
%and either
%\begin{equation}
%\frac{p(n)\cdot a(n)\cdot 2^j}{n}\to c \in (0,1)\quad \textnormal{or}\quad
%    \infty \leftarrow p(n) = o\Big(\sqrt{\frac{n}{a(n)}}\Big).
%\end{equation}
%
%\medskip
%
%\noindent {\sc Assumption $(A5)$}: For the random matrix ${\mathbf P}(n)$ as in \eqref{e:Y=PX},
%
%\begin{equation}\label{e:constraints_on_coordinate_matrix}
%    \frac{\log \sigma_1(\mathbf{P}(n))}{\log a(n)} \stackrel{\bbP}\to 0,\quad \frac{\log \sigma_p(\mathbf{P}(n))}{\log a(n)} \stackrel{\bbP}\to 0, \quad n \rightarrow \infty.
%\end{equation}
%
%
%\medskip
\subsection{Spectral graph theory}
We now recap some definitions of graphs and spectral graph theory that will be used in Section \ref{s:main_results}.

\begin{definition}\label{d:graph} A \textit{graph} $G$ is an ordered pair $(E,V)$, where $V$ denotes a set of \textit{vertices} and $E\subseteq \{e_{x,y}: x,y \in V, \hspace{0.5mm} x \neq y \}$ represents a set of \textit{edges}. Each $e_{x,y}$ is an \textit{undirected edge} ($e_{x,y} \equiv e_{y,x}$) connecting the pair of vertices $x$ and $y$. A graph $G = (E,V)$ is called a \textit{disjoint union of graphs} $G_1,\hdots,G_k$ if $V_i \cap V_j = \varnothing$ and $E_j \cap E_j = \varnothing$ for $1\leq i,j\leq k$ and $ i \neq j$. In this case, we write $E = \bigcupdot_i E_i$ and $V = \bigcupdot_i V_i$.
\end{definition}

\begin{definition}\label{d:simple_graph} A graph $G$ is called \textit{simple} if, for every $x,y \in V$, there exist $k$ and a \textit{path} (sequence of vertices) $\{v_i\}_{1\in \{1,\hdots,k\}} \subseteq V$ such that $\{e_{x,v_1},e_{v_1,v_2},\hdots,e_{v_{k-1},v_k},e_{v_k,y}\} \subseteq E$. A graph $G$ is called \textit{complete} if for every $x,y \in V$, $x \neq y$, $e_{x,y} \in E$.
\end{definition}

%\label{d:complete_graph}

%\begin{definition}\label{d:complete_graph_union} A graph $G = (E,V)$ is a \textit{disjoint union of graphs} $G_1,\hdots,G_k$ if $V_i \cap V_j = \varnothing$ and $E_j \cap E_j = \varnothing$ for $1\leq i,j\leq k$ and $ i \neq j$. In addition, $E = \bigcup E_i$ and $V = \bigcup V_i$.
%\end{definition}

\begin{definition}\label{d:adjacency_mat} Let $G$ be a graph with $|V| = p$, where the vertices are enumerated as integers $1,\hdots,n$. Then, the \textit{adjacency matrix} associated with $G$ is defined as the symmetric matrix given by
\begin{equation}\label{e:A(G)}
{\mathbf A}(G) := [ \mathds{1}_{ \{e_{i,j} \in E \}}]_{1\leq i \leq j \leq p}.
\end{equation}
In particular, each entry of ${\mathbf  A}(G)$ is either zero or one. In turn, the \textit{degree matrix} ${\mathbf  D}(G)$ of the graph $G$ is the diagonal matrix with entries ${\mathbf  D}(G)_{ii} := \sum_{j=1}^p {\mathbf  A}(G)_{ij}$, $i = 1,\hdots,p$. Then, the \textit{graph Laplacian (matrix)} of $G$ is defined as the difference ${\mathbf  L}(G) := {\mathbf  D}(G) - {\mathbf  A}(G)$.
\end{definition}

\begin{example}
Let $G$ be a complete graph containing $p$ vertices. Then,
\begin{equation}\label{e:A(G)=11^T-I}
{\mathbf  A}(G) = {\mathbf 1}{\mathbf 1}^\top - I,
\end{equation}
where ${\mathbf 1}^\top = [1,\hdots,1] \in \bbR^p$.
\end{example}

%\begin{definition}\label{d:graph_laplacian}
%    Let $G$ be a graph with $|V| = n$, where the vertices are enumerated as integers $1,\hdots,n$. Consider the adjacency matrix $A(G)$. Then, the \textit{graph Laplacian} is defined as
%    $$
%    L(G) := D(G) - A(G),
%    $$
%    where $D(G)$ is a diagonal matrix with entries
%    $
%    D(G)_{ii} := \sum_{j=1}^nA(G)_{ij}.
%    $
%    We call $D(G)$ the \textit{degree matrix} of the graph $G$
%\end{definition}
We now define the construction of a graph based on a collection of points in $\mathbb{R}^d$ and a distance threshold $\varepsilon > 0$.
\begin{definition}\label{d:epsilon_threshold} Let $\mathcal{X} = \{\x_1,\hdots,\x_p\} \subseteq \mathbb{R}^d$ be some collection of $p$ points.  For a fixed threshold $\varepsilon > 0$, we define $G_{\varepsilon}(\mathcal{X}) =(V,E)$ to be the \textit{graph induced by an $\varepsilon$--threshold}  where
$V =\mathcal{X}$ and $E=\{ e_{\x_i,\x_j} :  \| \x_{i} - \x_{j}\| < \varepsilon, \ i \neq j \}$. That is, $G_{\varepsilon}(\mathcal{X})$ is the graph obtained by connecting points that are within a distance $\varepsilon$ of one another. In particular, $G_{\varepsilon}(\mathcal{X})$ has adjacency matrix
\begin{equation}\label{e:adjacency_mat}
{\mathbf  A}\big(G_{\varepsilon }(\mathcal{X})\big) :=  \big[ \mathds{1}_{ \{ \| \x_{i} - \x_{j} \| < \varepsilon , \ i \neq j \}} \big]_{1 \leq i  \leq j \leq p}.
\end{equation}
and graph Laplacian
\begin{equation}\label{e:graph_laplacian_mat}
{\mathbf  L}\big(G_{\varepsilon }(\mathcal{X})\big)  :=  {\mathbf  D}\big(G_\varepsilon(\mathcal{X})\big) - {\mathbf  A}\big(G_{\varepsilon }(\mathcal{X})\big).
\end{equation}
For simplicity of notation, we write ${\mathbf  A}_{\varepsilon }(\mathcal{X}) = {\mathbf  A}\big(G_{\varepsilon }(\mathcal{X})\big) $ and ${\mathbf  L}_{\varepsilon }(\mathcal{X}) = {\mathbf  L}\big(G_{\varepsilon }(\mathcal{X})\big) $.
\end{definition}

\section{Main result}\label{s:main_results}

Consider the collection of statistics
\begin{equation}\label{e:scaling_estimate}
\widehat{\mathsf{H}}_p := \{ \widehat{H}_1,\hdots,\widehat{H}_p\}\quad\textnormal{with}\quad\widehat{H}_\ell := \frac{\log \lambda_\ell\big( \boldsymbol{{\mathcal W}}(a(n)2^j) \big)}{2 \log a(n) }-\frac{1}{2},\quad \ell \in \{1,\hdots, p\}.
\end{equation}
For terminological simplicity, hereinafter we refer to the rescaled and shifted wavelet log-eigenvalues $\widehat{H}_\ell$ as in \eqref{e:scaling_estimate} simply as ``wavelet log-eigenvalues". Heuristically, we can interpret each $\widehat{H}_\ell$ as an ``estimator'' of the corresponding ordered Hurst exponent $H_{(\ell)}$ (\textbf{n.b.}: the latter stems from realizations of the random variable ${\mathcal H}$). Based on recent results from random matrix theory, the collection of estimators $\widehat{\mathsf{H}}_p$ coalesce around the Hurst modes $\breve{H}_{1},\hdots, \breve{H}_{r}$ in the three-way limit \eqref{e:p(n),a(n)_conditions}, with increasing probability (see Theorem \ref{t:main_theorem_discrete} for the precise statement). As alluded to in the Introduction, in this section we leverage this fact and propose the \textbf{WRMSM} algorithm (\textbf{W}avelet \textbf{R}andom \textbf{M}atrix, modified \textbf{S}pectral clustering and \textbf{M}odel selection) for the estimation of the Hurst distribution $\pi(dH)$. With the purpose of defining the algorithm, first consider the following definitions.
    \begin{definition}\label{d:within_sum_of_squares}
    Fix $n \in \Naturals$. A \textit{clustering scheme} for $\widehat{\mathsf{H}}_p$ as in \eqref{e:scaling_estimate} is any finite collection of sets 
    \begin{equation}\label{e:mathcal C_def}
    \mathcal{C} = \{C_1,\hdots,C_{\widehat{r}}\}
    \end{equation}
    satisfying  
    \begin{equation}\label{e:r-hat}
    \bbN \ni \widehat{r} :=  \# \textnormal{ clusters in }{\mathcal C}  \quad \textnormal{and}\quad \widehat{\mathsf{H}}_p = \bigcupdot^{\widehat{r}}_{k=1}C_k.
    \end{equation}
    Fix any ${\mathcal C}$ as in \eqref{e:mathcal C_def}. For each $j=1,\hdots,\widehat{r}$, we naturally define the \textit{estimators of the Hurst modes and their probabilities under $\pi(dH)$}, respectively, as
    \begin{equation}\label{e:H-breve-hat_pi-hat}
    \widehat{\breve{H}}_j = \textnormal{mean}(C_j) = \frac{1}{|C_j|} \sum_{\ell: \widehat{H}_{\ell} \in C_j} \widehat{H}_{\ell} \quad  \textnormal{and} \quad \widehat{\pi}_j= \frac{|C_j|}{p}.
    \end{equation}
    In addition, the associated \textit{intra-cluster standard deviation} is given by
    \begin{equation}\label{e:ICSD}
        \textnormal{ICSD}(\mathcal{C}) = \sum^{\widehat{r}}_{j=1} \Big(\frac{1}{|C_j|}\sum_{\ell: \widehat{H}_{\ell} \in C_j} (\widehat{H}_{\ell} - \widehat{\breve{H}}_j )^2\Big)^{1/2}.
    \end{equation}
\end{definition}

In Definition \ref{d:within_sum_of_squares}, it should be stressed that all natural estimators $\widehat{r}$, $(\widehat{\breve{H}}_j, \widehat{\pi}_j)_{j = 1,\hdots, \widehat{ r}}$, are defined \textit{given} the clustering scheme ${\mathcal C}$. In this context, it is clear that the construction of an estimation algorithm for $\pi(dH)$ further requires a clustering methodology. 

%Structurally speaking, the algorithm contains two main components: $(i)$ given a precision hyperparameter $\varepsilon > 0$, an estimation subroutine that breaks up $\widehat{\mathsf{H}}_p$ into a clustering scheme ${\mathcal C}_{\varepsilon} = \{{\mathcal C}_{1,\varepsilon}, \hdots, {\mathcal C}_{\widehat{r}_{\varepsilon},\varepsilon}\}$, where $\widehat{r}_{\varepsilon}$ is an estimation of $r$ as in \eqref{e:pi(dh)}; $(ii)$ a built-in model selection procedure to pick a ``good" precision hyperparameter $\varepsilon$.  

We are now in a position to define \textbf{WRMSM} as well as the \textit{Hurst distribution $\varepsilon$-precision estimation subroutine} (\textbf{HD$\varepsilon$ES}) that \textbf{WRMSM} builds upon. For expository purposes, we present them sequentially in the form of pseudocode, and then provide some interpretation. In the description of \textbf{WRMSM}, note that \textbf{HD$\varepsilon$ES} depends on a precision hyperparameter $\varepsilon > 0$. 

%%%%%%%%%%%%%%%%%%%%%%%%%%%%%%%%%%%%%%%%%%%%%%%%%%%%%%%%%%%%%%%%%%%%%%%%%%
%\vspace{-0.5cm}
{\small
\begin{center}
\begin{tabular}{|l|}
\hline \\
\multicolumn{1}{|c|}{pseudocode for the \textbf{WRMSM} algorithm}\\ 
\\
\hline\\
\textbf{Input}: $M > 0$, $m \in \bbN$, $\widehat{\mathsf{H}}_p$ as in \eqref{e:scaling_estimate}.\\
\\
\textbf{Step 1}: For $k = 1\hdots,m$, let $\varepsilon_k = \frac{k \cdot M}{m}$ and apply \textbf{HD$\varepsilon$ES} to obtain ${\mathcal C}_{\varepsilon_k}$.\\
\\
\textbf{Step 2}: Pick ${\mathcal C}_{\varepsilon_{ms}}$, where %\vspace{0.1cm}
\parbox[t]{8cm}{
\begin{equation}\label{e:model_selection_threhsold}
\hspace{-5cm}\varepsilon_{ms}:= \textnormal{argmin}_{k=1,\hdots,m} \textnormal{ICSD}( \mathcal{C}_{\varepsilon_k}).
\end{equation}}\\

\textbf{Output}:  $\widehat{r}_{\varepsilon_{ms}}$, $(\widehat{\breve{H}}_j ,  \widehat{\pi}_j)_{j=1,\hdots,\widehat{r}_{\varepsilon_{ms}}}$ (see \eqref{e:r-epsilon_def}, \eqref{e:r-hat} and \eqref{e:H-breve-hat_pi-hat}) associated with ${\mathcal C}_{\varepsilon_{ms}}$.\\ \\
\hline
\end{tabular}
\end{center}
}

\vspace{2mm}

{\small
\begin{center}
\begin{tabular}{|l|}
\hline \\
\multicolumn{1}{|c|}{pseudocode for \textbf{HD$\varepsilon$ES}}\\ 
\\
\hline\\
\textbf{Input}: $\varepsilon > 0$, $\widehat{\mathsf{H}}_p$ as in \eqref{e:scaling_estimate}.\\
\\
\textbf{Step 1}: Construct a graph induced by the $\varepsilon$-threshold as in Definition $\ref{d:epsilon_threshold}$.\\
\\
%\textbf{Step 2}: Let ${\mathbf A}_\varepsilon(\widehat{\mathsf{H}}_p)$ be the associated adjacency matrix as in  $\eqref{e:adjacency_mat}$;\\
%\\
%\textbf{Step 2}: Compute the Laplacian ${\mathbf L}_\varepsilon(\widehat{\mathsf{H}}_p)$ as in $\eqref{e:graph_laplacian_mat}$;\\ 
%\\
\textbf{Step 2}: Compute the eigenvalues $\{ \theta_\ell \}_{1\leq \ell \leq p}$ (ordered increasingly) as well as the corresponding \\
eigenvectors $\{\mathbf{u}_\ell\}_{1\leq \ell \leq p}$ of the Laplacian ${\mathbf L}_\varepsilon(\widehat{\mathsf{H}}_p)$ as in $\eqref{e:graph_laplacian_mat}$.\\
\\
\textbf{Step 3}: Let \\ 
\parbox[t]{14cm}{
\begin{equation}\label{e:r-epsilon_def} \widehat{r}_\varepsilon = \textnormal{argmax}_{1\leq \ell \leq p}|\theta_{\ell+1}-\theta_{\ell} |.
\end{equation}}\\
\textbf{Step 4}: Let ${\mathbf U} \in \bbR^{p\times \widehat{r}_\varepsilon}$ be the matrix with $\mathbf{u}_1,...,\mathbf{u}_{\widehat{r}_\varepsilon}$ as columns.\\
\\
\textbf{Step 5}: For $i = 1,...,p$, let $\mathbf{y}_i \in \bbR^{\widehat{r}_{\varepsilon}}$ be the vector corresponding to the $i-$th row of ${\mathbf U}$.\\ 
\\
\textbf{Step 6}: Organize the points $\{\mathbf{y}_i\}_{1\leq i \leq p}$  with the $k$-means algorithm into clusters $C'_1,...,C'_{\widehat{r}_\varepsilon}$.\\ 
\\
%\textbf{Step 9}: For each $j=1,\hdots,\widehat{r}_\varepsilon$, let $\widehat{\breve{H}}_j = \textnormal{mean}(C_j)$ and $\widehat{\pi}_j=|C_j|/p$;\\ 
%\\
\textbf{Output}:  $\widehat{r}_{\varepsilon}$, $(\widehat{\breve{H}}_j ,  \widehat{\pi}_j)_{j=1,\hdots,\widehat{r}_{\varepsilon}}$ (see \eqref{e:r-epsilon_def}, \eqref{e:r-hat} and \eqref{e:H-breve-hat_pi-hat}) associated with ${\mathcal C}_{\varepsilon}$.\\ \\
\hline
\end{tabular}
\end{center}
}

\vspace{2mm}

In summary, \textbf{WRMSM} works as follows. For multiple choices of a precision hyperparameter $\varepsilon > 0$, \textbf{HD$\varepsilon$ES} breaks up $\widehat{\mathsf{H}}_p$ into a clustering scheme ${\mathcal C}_{\varepsilon}$. Intuitively speaking, the clustering scheme ${\mathcal C}_{\varepsilon}$ is ``good" if and only if $\textnormal{ICSD}(\mathcal{C}_{\varepsilon}) \approx 0$ (this reasoning is rigorously established in Lemma \ref{p:ICDS_to_zero}). For this to happen, ${\mathcal C}_{\varepsilon}$ must mimic the distribution $\pi(dH)$ in the sense that $\widehat{r}_{\varepsilon} \approx r$, $\widehat{\breve{H}}_j \approx \breve{H}_j$ and $\widehat{\pi}_j \approx \pi_j$. Then, a built-in \textit{model selection} procedure in \textbf{WRMSM} proceeds to pick the ``best choice" of precision hyperparameter $\varepsilon$ by minimizing the ICSD (see \eqref{e:model_selection_threhsold}). Even though any ``good" choice of $\varepsilon$ leads to consistent estimation by means of \textbf{HD$\varepsilon$ES} (see Proposition \ref{t:fixed_epsilon_consistency}), picking the ``best one" ensures the finite-sample performance of  \textbf{WRMSM} is greatly improved.

In turn, \textbf{HD$\varepsilon$ES} is a subroutine that builds upon the spectral clustering technique as outlined in von Luxburg \cite{vonLuxburg:2007} (on the use of $k$-means in \textbf{HD$\varepsilon$ES}, see Remark \ref{r:rates_of_convergence}, ($iii$)). Notably, \textbf{HD$\varepsilon$ES} does \textit{not} require prior knowledge of the number of clusters. Instead, the number of clusters is determined by the estimator of the number of modes $r$ (see \eqref{e:r-epsilon_def}). 
%In particular, for good choices of the precision hyperparameter $\varepsilon$, \textbf{HD$\varepsilon$ES} provides consistent estimation of the Hurst distribution. This is established in Theorem \ref{t:fixed_epsilon_consistency} below.

Even though it inherently addresses the issue of how to pick the precision hyperparameter $\varepsilon$ in $\textbf{HD$\varepsilon$ES}$, \textbf{WRMSM} itself requires two hyperparameters $M>0$ and $m \in \bbN$ that define the search space ${\mathcal E}$. However, $M$ and $m$ are much easier to pick in practice and do not require direct knowledge of some aspect of $\pi(dH)$ (\textbf{n.b.}: by contrast, $\varepsilon$ is directly related to $\Delta_{\min}$ as in \eqref{e:threshold_0}; see Remark \ref{r:rates_of_convergence}, ($i$)). The parameter $M$ simply acts as an upper bound on the set of possible values $\varepsilon$ to check, whereas the parameter $m$ controls how finely we would like to search the interval $(0,M)$ for parameters (see, again, Remark \ref{r:rates_of_convergence}, ($i$)).

The following theorem is the main mathematical result of the paper, i.e., it demonstrates the consistency of \textbf{WRMSM} in the estimation of the Hurst distribution $\pi(dH)$. Note that the theorem requires the existence of a ``good" choice of precision hyperparameter $\varepsilon_g \in (0, \Delta_{\min})$, where
\begin{equation}\label{e:threshold_0}
\Delta_{\min} := \left\{\begin{array}{cc}
 \textnormal{min}_{i\neq j}|\breve{H}_i - \breve{H}_j|, & \textnormal{if }r > 1;\\
 \infty, & \textnormal{if }r = 1.
\end{array}\right.
\end{equation}
%for $r>1$ with $\{\breve{H}_1,\hdots,\breve{H}_r \}$ as in assumption ($A1$). Note, in the case that $r=1$ (unimodal distribution), the following theorems hold for all choices of $\Delta_{\min}>0$.

\begin{theorem}\label{t:WRMSM_consistency}
Suppose assumptions ($A1-A5$) and ($W1-W3$) hold. Let $\widehat{\mathsf{H}}_p$ be as in \eqref{e:scaling_estimate}, and let $\Delta_{\min}$ be as in \eqref{e:threshold_0}. Pick $M>0$ and $m \in \bbN$ such that, for
\begin{equation}\label{e:E=choice_set}
\mathcal{E} =\Big\{ \frac{k\cdot M}{m}\ : \ k = 1,\hdots m\Big\} \backslash \{\Delta_{\min}\},
\end{equation}
 there is at least one $ \varepsilon_g \in \mathcal{E} $  satisfying
    \begin{equation}\label{e:there_is_a_good_epsilon}
        \varepsilon_g < \Delta_{\min}.
    \end{equation}
Then, the \textbf{WRMSM} algorithm is consistent. Namely, for any $\eta >0$, as $n\to\infty$,
    \begin{equation}\label{e:good_epsilon_consistency_ms}
    \bbP\Big( \{ \widehat{r}_{\varepsilon_{ms}} = r\} \bigcap \Big\{  \max_{i = 1,\hdots, r}| \widehat{\breve{H}}_i -\breve{H}_i| < \eta \Big\}  \bigcap \Big\{  \max_{i = 1,\hdots, r}| \widehat{\pi}_i -\pi(\breve{H}_i)| < \eta \Big\} \Big) \to 1.
    \end{equation}
\end{theorem}

It is interesting to note that, beyond consistency, relation \eqref{e:good_epsilon_consistency_ms} implies that, with probability going to 1 as $n \rightarrow \infty$,  all estimators $\widehat{\breve{H}}_i$  and $\widehat{\pi}_i$, $i = 1,\hdots,r$, lie at \textit{uniformly} small distances from their target parameters.

%\begin{remark}\label{r:why_not_k_means}
% In this paper, we aim to establish a provably consistent algorithm for the estimation of the Hurst distribution. For the sake of illustration, note that there are some significant technical difficulties involved in using the popular $k$-means algorithm directly in place of \textbf{Step 1}. First off, $k$-means assumes $r$ is known. However, this is not true in modeling practice starting from measurements of the form \eqref{e:Y(ell)=P(n)X(ell)_intro}. Second, the bounds constructed in Abry et al.\ \cite{abry:didier:orejola:wendt:2024} on the relative behavior of the scaling estimates \eqref{e:scaling_estimate} are not sufficiently tight \GDcomment{This is unclear}. This makes analysis difficult as $k$-means only makes use of relative distances between observations. By contrast, in the context of \textbf{Step 1}, the bounds are, indeed, sufficient in establishing consistency. 
%\end{remark}

\begin{remark} \label{r:rates_of_convergence}\textit{ }

\begin{itemize}
\item [$(i)$] The condition on the existence of some $\varepsilon_g$ as in \eqref{e:there_is_a_good_epsilon} is essential for the consistency of \textbf{WRMSM}. Intuitively, if the algorithmic search is only conducted over values $\varepsilon > \Delta_{\min}$, then there are two modes $\breve{H}_{\ell}$ and $\breve{H}_{\ell'}$ close enough to each other ($|\breve{H}_{\ell} -\breve{H}_{\ell'} | = \Delta_{\min}$) that cannot be ``resolved" by the algorithm (see Lemma \ref{p:ICDS_to_zero} for a rigorous expression of this statement).  

In turn, it is clear that the existence of $\varepsilon_g \in {\mathcal E}$ satisfying \eqref{e:there_is_a_good_epsilon} depends on the choices of $M$ and $m$. In the simulation studies appearing in Section \ref{s:computational_studies}, we make a heuristic choice for $M$ as
\begin{equation}\label{e:model_selection_threshold}
    M = \frac{1}{2 \log a(n)}\log\Big(\frac{\lambda_p(\boldsymbol{\mathcal{W}}(2^j))}{\lambda_1(\boldsymbol{\mathcal{W}}(2^j))} \Big).
\end{equation}
From Lemma \ref{p:log_eigenvalues_count} (see expressions \eqref{e:log_eigenvalue_bound1}--\eqref{e:G_upper}), it follows that this choice of $M$ is a modest upper bound for a ``good" value $\varepsilon$  to perform spectral clustering with. In turn, $m$ should be chosen not too large. More precisely, if the hyperparameter value $\varepsilon_1 = M/m$ is significantly smaller than $1/p$, then one may end up with ${\mathcal C}_{\varepsilon_{ms}}$ containing single-point clusters. This is undesirable because the inclusion of single-point clusters is a trivial way of minimizing the ICSD. %Also, the precision hyperparameter $\varepsilon_{ms}$ output by \textbf{WRMSM} is often the largest $\varepsilon \in\mathcal{E}$ such that $\varepsilon  < \Delta_{\min}$.

\item [$(ii)$]
In the face of \textbf{WRMSM}'s consistency as stated in Theorem \ref{t:WRMSM_consistency}, it is natural to ask if one can additionally provide rates of convergence or confidence bounds. 

It is clear that expression \eqref{e:log_eigenvalue_bound1} plays a central role in the overall construction of the proof of Theorem \ref{t:WRMSM_consistency}. Hence, in order to obtain rates of convergence it suffices to establish the joint asymptotic distribution of the extreme eigenvalues of the fixed-scale wavelet random matrix ${\boldsymbol {\mathcal W}}(2^j) $. This, in turn, is an interesting open problem that calls for future efforts (cf.\ Bai, Miao and Pan \cite{bai:miao:pan:2007}, Bai, Liu and Wong \cite{bai:liu:wong:2011}, Paul \cite{paul:2012}, Ding and Yang \cite{ding:yang:2020}, Bao et al.\ \cite{bao:lee:xu:2024}). 

\item [$(iii)$]  %\begin{remark}\label{r:on_Step_1}
Observe that, in the construction of \textbf{HD$\varepsilon$ES}, we make use of the well-known \textit{$k$-means algorithm}, which is recapped in Section \ref{s:k-means}. Briefly, $k$-means aims to partition a dataset into $\kappa$ clusters by iteratively assigning data points to the nearest cluster center, and then by updating the cluster centers based on the means of the assigned points. Note that the $k$-means algorithm assumes the number of clusters $\kappa$ is \textit{known}. Also, it often yields different clustering schemes depending on initialization (see Pe$\tilde{\textnormal{n}}$a et al.\ \cite{pena:lozano:Larranaga:1999}), which may be an issue in the analysis of consistency. 

Nevertheless, in the context of \textbf{HD$\varepsilon$ES}, $k$-means is fed the \textit{estimated} number of clusters $\widehat{r}_{\varepsilon}$. Furthermore, after $k$-means clusters the data points $\{\mathbf{y}_i\}_{i = 1,\hdots, p}$, we can proceed in the analysis of consistency by making use of the property described in Lemma \ref{l:k-means_converges}. This analysis appears in the proof of Proposition \ref{t:fixed_epsilon_consistency}.
%\end{remark}
\end{itemize}
\end{remark}

\section{Computational studies}\label{s:computational_studies}

%\GDcomment{I corrected a number of typos, some of which go unmarked. The section generally reads well. One issue that remains is, like through the rest of the paper, the use of the expression ``spectral clustering'' to refer to our algorithm.}

To assess the practical relevance of the theoretical results stated in Section \ref{s:main_results}, we make use of Monte Carlo experiments based on 1000 independent realizations of $p$-variate measurements, where ${\mathbf P} = {\mathbf P}(n)$ is a randomly chosen orthogonal matrix and $X(t)$ as in \eqref{e:Y(ell)=P(n)X(ell)_intro} is made up of independent univariate fBms with $n = 2^{14}$ (such sample sizes are realistic, for example, in the context of the analysis of infraslow brain activity: see LaRocca et al.\ \cite{LaRocca:Wendt:vanWassenhove:Ciuciu:Abry:2021}). The univariate fBms are generated using the \textbf{R} package \texttt{somebm} \cite{somebm}. The $p\times p$ diagonal matrix $\mathds{H}_n$ (see \eqref{e:mathds_hn}) is obtained by drawing $p$ i.i.d.\ samples from a discrete distribution $\pi(dH)$ for each realization independently. The wavelet transformation is generated by means of Mallat's algorithm based on a Daubechies filter with $N_\psi = 2$. In particular, to compute the wavelet transformation we make use of the \textbf{R} package \texttt{wavelets} (see Aldrich \cite{wavelets}).  Throughout this computation study, we use $n = 2^{14}$, $p = 2^6$, $a(n) = 2^4$, and $j = 1$. When performing model selection, we make use of the heuristic choice of $M$ as in \eqref{e:model_selection_threshold} and set $m=10$.

\begin{figure}
    \centering
    \includegraphics[scale=0.4]{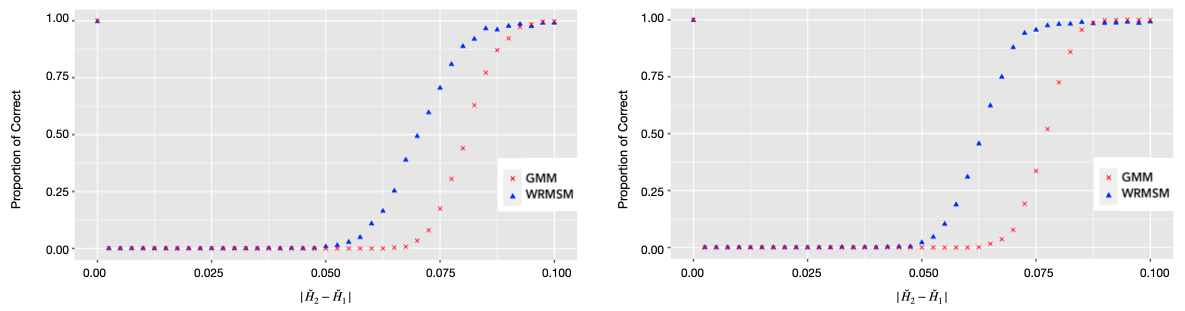}
 \caption{ \textbf{Bimodal Hurst distributions.} Proportion of correct identification of the number of Hurst modes ($\widehat{r}_{\varepsilon_{ms}}=2$) over 1000 Monte Carlo runs. \textbf{Left plot}: $\pi(dH)$ non-uniform distribution. \textbf{Right plot}: $\pi(dH)$ uniform distribution. In both plots, \textbf{WRMSM} and \textbf{GMM}-based clustering appear in blue and red, respectively.} 
    \label{fig:2_Hurst_prop_epsilon}
\end{figure}

To study the performance of the \textbf{WRMSM} algorithm, we consider a variety of Hurst distributions $\pi(dH)$. For different choices of $\pi(dH)$, we test the accuracy of the method by computing the proportion of times the estimated number of clusters,  $\widehat{r}_{\varepsilon_{ms}}$, equals $r = |\textnormal{supp}\ \pi(dH)|$.

To gauge the performance of \textbf{WRMSM}'s performance, we use the celebrated Gaussian mixture model-based clustering (\textbf{GMM}; see Fraley and Raftery \cite{Fraley:Raftery:2002}). \textbf{GMM}-based clustering makes use of Gaussian mixture models assuming unequal variances. Models are estimated by means of the EM algorithm (Expectation-Maximization) initialized by hierarchical model-based agglomerative clustering. The optimal model is then selected according to BIC. The \textbf{R} package \texttt{mclust} \cite{mclust} was used to implement \textbf{GMM}-based clustering. We compare \textbf{WRMSM}'s with \textbf{GMM}'s ability to recover $r = |\textnormal{supp}\ \pi(dH)|$.

For the (bimodal) case where $\textnormal{supp}\ \pi(dH) = \{\breve{H_1},\breve{H_2}\}$, we consider varying the values $\breve{H}_1=0.25$, $\breve{H}_2=0.25+\Delta$, as a function of $\Delta \in [0.0,0.10]$. The results are reported in Figure~\ref{fig:2_Hurst_prop_epsilon} (right). They show that, for $\pi(dH) \stackrel{d}= \textnormal{Unif}\{\breve{H_1},\breve{H_2}\}$, \textbf{WRMSM}-based estimation begins to correctly identify two modes near $\Delta = 0.05$. Moreover, it identifies two modes with probability greater than $3/4$ when $\Delta \geq 0.075$. This stands in contrast to \textbf{GMM}'s capabilities. That is, \textbf{GMM} begins to correctly identify two modes near $\Delta = 0.07$. It further identifies two modes with probability greater than $3/4$ when $\Delta \geq 0.085$. Similar results are observed for non-uniform Hurst distributions, e.g., $\pi(\breve{H_1}) = 1/3$ and  $\pi(\breve{H_2}) = 2/3$ (see Figure~\ref{fig:2_Hurst_prop_epsilon}, left). Note that both clustering methods are capable of identifying the unimodal case properly.

\begin{figure}[ht]
    \centering
    \includegraphics[scale=0.4]{"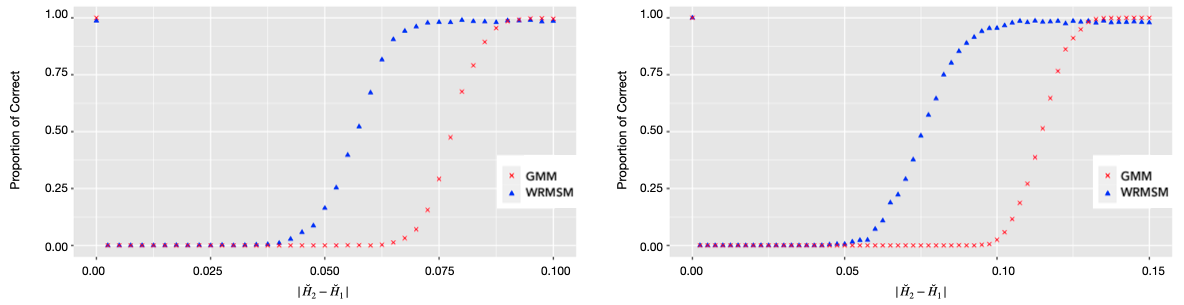"}
 \caption{ \textbf{Trimodal Hurst distributions.} Proportion of correct identification of the number of Hurst modes ($\widehat{r}_{\varepsilon_{ms}}=3$) over 1000 Monte Carlo runs. \textbf{Left plot}: $\breve{H}_1, \breve{H}_3$ fixed with $\breve{H}_2$ varying. \textbf{Right plot}: $\breve{H}_1, \breve{H}_2, \breve{H}_3$ equidistant. In both plots, \textbf{WRMSM} and \textbf{GMM}-based clustering appear in blue and red, respectively.}
    \label{fig:TEST_3_Hursts}
\end{figure}

The results for the trimodal distribution ($\textnormal{supp}\ \pi(dH) = \{\breve{H_1},\breve{H_2},\breve{H_2}\}$) are reported in Figure ~\ref{fig:TEST_3_Hursts}. There, we consider two distinct varying arrangements of $\breve{H}_1,\breve{H}_2,\breve{H}_3$ with $\pi(dH) \stackrel{d}= \textnormal{Unif}\{\breve{H}_1,\breve{H}_2,\breve{H}_3\}$.
The results for $ \breve{H}_1 = 0.25$, $\breve{H}_2=0.25+\Delta$, $\breve{H}_3 = 0.7$, $\Delta \in [0,0.1]$, are reported in Figure~\ref{fig:TEST_3_Hursts} (left). Also, for  $ \breve{H}_1 = 0.5-\Delta$, $\breve{H}_2=0.5$, $\breve{H}_3 = 0.5+\Delta$, $\Delta \in [0,0.1]$, the results appear in Figure~\ref{fig:TEST_3_Hursts} (right). In both cases, we observe superior performance of the \textbf{WRMSM} algorithm to that of \textbf{GMM}.

\begin{figure}[t]
    \centering
    \includegraphics[scale=0.4]{"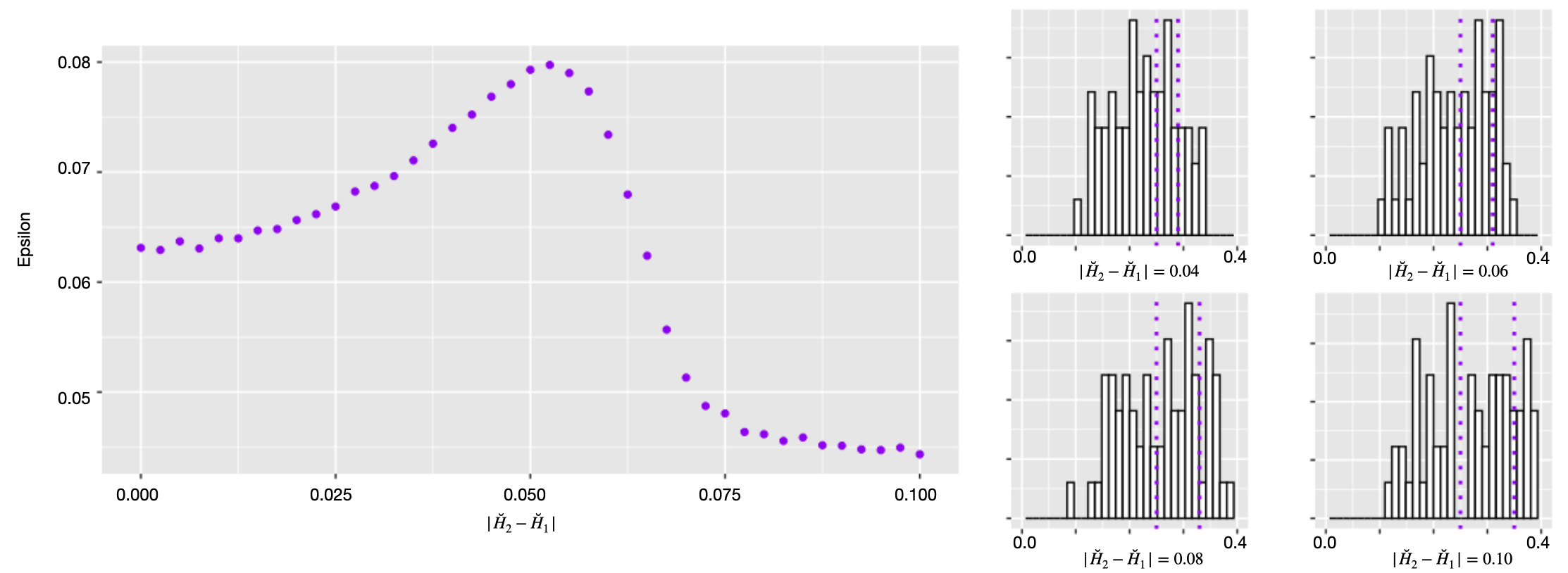"}
 \caption{ \textbf{Optimal $\varepsilon>0$ chosen via model selection (left plot).} The optimal hyperparameter value $\varepsilon_{ms}$ (see \eqref{e:model_selection_threhsold}) was computed by means of averaging over 1000 Monte Carlo experiments. \textbf{Instances of wavelet log-e.s.d.\ (right plots).} $|\breve{H}_2-\breve{H}_1| = 0.04$\textbf{ (top left)}, $|\breve{H}_2-\breve{H}_1| = 0.06$ \textbf{(top right)}, $|\breve{H}_2-\breve{H}_1| = 0.08$ \textbf{(bottom left)}, $|\breve{H}_2-\breve{H}_1| = 0.10$ \textbf{(bottom right)}.
 }
\label{fig:hist_epsilons}
\end{figure}

The analysis of the choice of $\varepsilon>0$ via \textbf{WRMSM} can be found in Figure~\ref{fig:hist_epsilons} (left). There we consider $\pi(dH) \stackrel{d}= \textnormal{Unif}\{\breve{H_1},\breve{H_2}\}$, for choices $\breve{H}_1=0.25$ and $\breve{H}_2 = 0.25 +\Delta$ with $\Delta \in [0,1]$. These results are consistent with Figure~\ref{fig:2_Hurst_prop_epsilon}. In particular, for $\Delta \in [0,0.05]$, we see that a typical instance of wavelet log-e.s.d.\ is visually unimodal (see top two histograms in Figure~\ref{fig:hist_epsilons}). In other words, unimodality is the “best” model, and a larger $\varepsilon > 0$ is necessary to capture this unimodality. In contrast, for $\Delta > 0.07$, we see that typical instances of the wavelet log-e.s.d.\ are bimodal, as observed in  Figure~\ref{fig:hist_epsilons} (bottom two histograms). Additionally, \textbf{WRMSM} identifies the correct number of modes, $\widehat{r}_{\varepsilon_{ms}}= 2$ (cf.\ Figure~\ref{fig:2_Hurst_prop_epsilon}, right). Accordingly, a smaller $\varepsilon > 0$ is chosen on average.
Overall, these findings confirm the finite-sample effectiveness of \textbf{WRMSM} as well as its satisfactory statistical performance.

\begin{remark}
Spectral clustering is known to be particularly effective for non-convex domains. By contrast, \textbf{GMM} is naturally suited for convex ones as a result of the general form of mixed-Gaussian distributions. In light of these facts, it is striking that \textbf{WRMSM} outperforms \textbf{GMM} since the convergence of wavelet log-e.s.d.\ data takes place in a convex domain.  
\end{remark}

\section{Application}\label{s:Application}

\begin{table}[htbp]
\centering
\begin{tabular}{|l|p{12cm}|}
\hline
\textbf{time series} & \textbf{description} \\
\hline
EXUSUKx & U.S./U.K.\ foreign exchange rate \\
\hline
TB3MS & 3-month treasury bill \\
\hline
WPSID62 & Producer price index by commodity: intermediate demand by commodity type (unprocessed goods for intermediate demand)\\
\hline
FEDFUNDS & Effective federal funds rate \\
\hline
OILPRICEx & Crude oil, spliced WTI and cushing \\
\hline
CPIAPPSL & CPI: apparel \\
\hline
AAA & Moody's seasoned AAA corporate bond yield \\
\hline
CPIMEDSL & CPI: medical care \\
\hline
GS10 & 10-year treasury rate \\
\hline
GS5 & 5-year treasury rate \\
\hline
DNDGRG3M086SBEA & Personal cons.\ exp: nondurable goods \\
\hline
WPSFD49502 & Producer price index by commodity: final demand by personal consumption goods (finished consumer goods) \\
\hline
BAA & Moody's seasoned BAA corporate bond yield \\
\hline
IPNCONGD & IP: nondurable consumer goods \\
\hline
\end{tabular}
\caption{FRED-MD: selected macroeconomic time series and their descriptions (McCracken and Ng \protect \cite{mccracken:ng:2015}).}
\label{tab:economic_signals}
\end{table}

In a very general sense, a multivariate stochastic process is said to be \textit{cointegrated} if there exists a linear combination of its entry-wise components that displays an order of integration (i.e., memory) that is lower than that of the entry-wise components themselves (e.g., Marinucci and Robinson \cite{marinucci:robinson:2001}, Robinson \cite{robinson:2008}, Nielsen and Frederiksen \cite{nielsen:frederiksen:2011}, Shimotsu \cite{shimotsu:2012}, Wang and Phillips \cite{wang:phillips:2023}). In the context of the high-dimensional model \eqref{e:Y(ell)=P(n)X(ell)_intro}, for any fixed $n$ and for most choices of matrix of coordinates ${\mathbf P}(n)$, cointegration can be interpreted to mean that $r > 1$, i.e., the existence of multiple Hurst modes.

As an application of the theoretical results in Section \ref{s:main_results}, we investigate the presence of the property of cointegration in the so-named Federal Reserve Economic Data of monthly frequency macroeconomic time series (FRED-MD; see McCracken and Ng \cite{mccracken:ng:2015}). FRED-MD is a large database maintained by the St.\ Louis Federal Reserve Bank and put together as a convenient starting point for empirical analysis that requires ``Big Data". 

\begin{figure}[t]
    \centering
  \includegraphics[scale=0.5]{"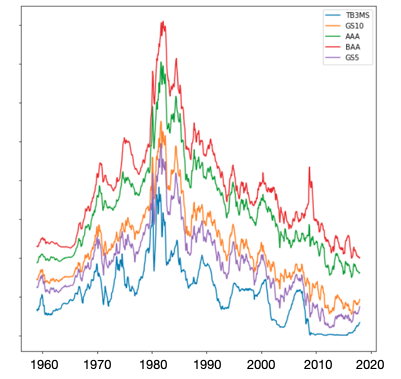"}  \includegraphics[scale=0.375]{"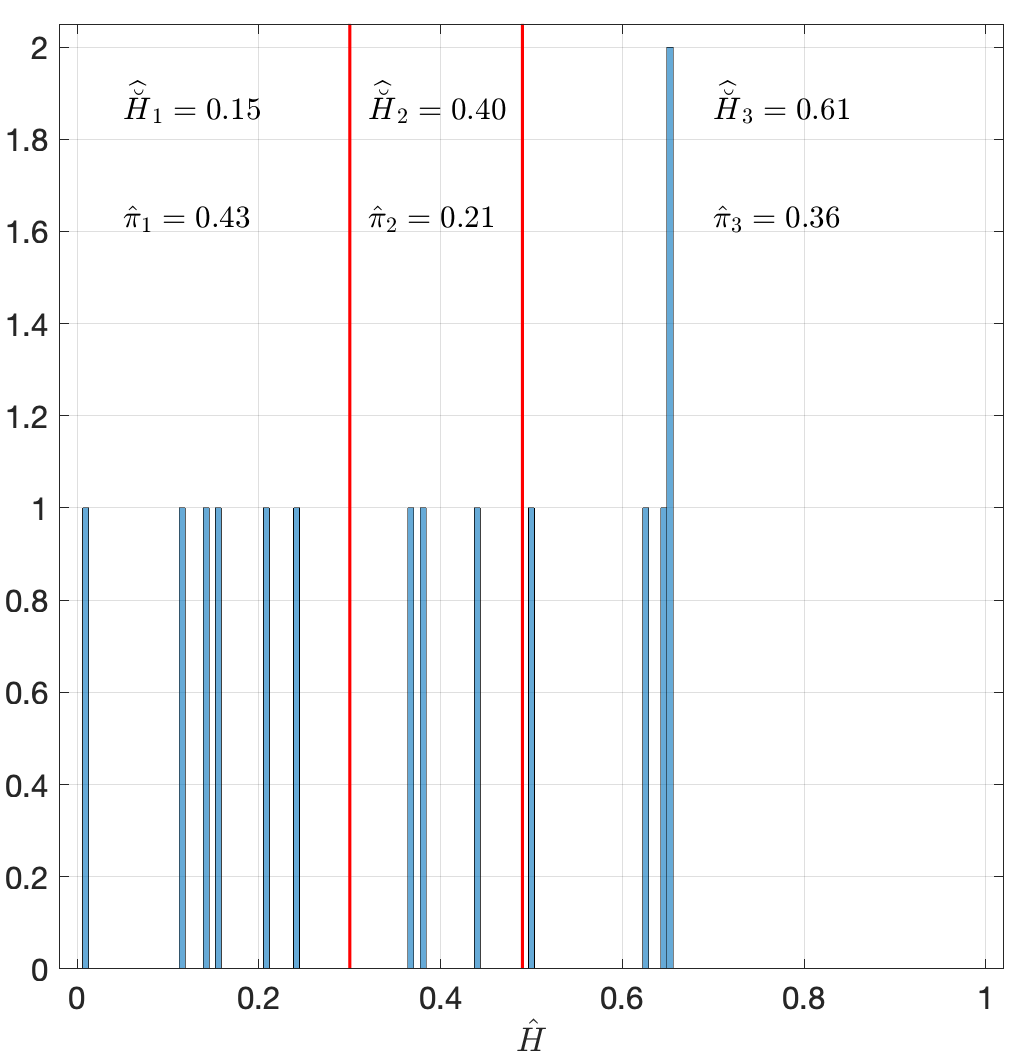"}
 \caption{ \textbf{ Left plot}: select FRED-MD economic indicators. The time series look both ``correlated" and nonstationary. \textbf{Right plot}: the vertical blue bars depict the histogram of the wavelet log-e.s.d.\ distribution. Wavelet log-eigenvalues weighted over multiple scales were used for debiased finite-sample convergence (cf.\ Abry and Didier~\protect\cite{abry:didier:2018:n-variate} and Abry et al.\ \protect\cite{abry:boniece:didier:wendt:2024}). The red lines mark the boundaries among the three clusters obtained from \textbf{WRMSM}. The estimated Hurst modes and their probabilities are also displayed.}
 
\label{fig:application_data}
\end{figure}

We apply \textbf{WRMSM} to a subset of $p = p(n) = 14$ time series of U.S.\ economic indicators appearing in FRED-MD with $n=709$ (monthly) measurements ranging from 1959 to 2018. The largest scale in consideration is $a(n)2^j=2^5$. See Table \ref{tab:economic_signals} for the economic indicators and their descriptions. The chosen individual time series were assessed for non-stationarity using two complementary statistical tests: the Augmented Dickey-Fuller (ADF) test and the Kwiatkowski-Phillips-Schmidt-Shin (KPSS) test. Additionally, these series were selected because certain pairs of series displayed evidence of (pairwise) cointegration (see Figure \ref{fig:application_data}, left). To better handle the different orders of magnitude of variance across data, each series was standardized by normalizing with respect to the standard deviation of the first differences. In addition, wavelet log-eigenvalues weighted over multiple scales ($j_1=2$ to $j_2=5$) were used for debiased finite-sample convergence (cf.\ Abry and Didier~\protect\cite{abry:didier:2018:n-variate} and Abry et al.\ \protect\cite{abry:boniece:didier:wendt:2024}). The method works equally as well at the fixed scale $2^j=2^5$. 

In Figure \ref{fig:application_data} (right), we see the wavelet log-e.s.d.\ formed from the selected time series. In our analysis, we select the parameter $M$ as in \eqref{e:model_selection_threshold}, and $m$ was fixed to be $10$. The \textbf{WRMSM} estimates for the number of modes, the modes themselves as well as their probabilities are given by, respectively, $\widehat{r}_{\varepsilon_{ms}}=3$, $(\widehat{\breve{H}}_1, \widehat{\breve{H}}_2, \widehat{\breve{H}}_3 ) =  (0.15, 0.40,0.61)$ and $(\widehat{\pi}_1, \widehat{\pi}_2, \widehat{\pi}_3 ) =(0.43, 0.21, 0.36)$. This provides evidence for the presence of cointegration among the selected $p = p(n) = 14$ time series from FRED-MD.

%\begin{remark}\label{r:signal_standardization}

%\end{remark}

\section{Conclusion and discussion}\label{s:conclusion}
In this paper, we bring together wavelet random matrices, spectral clustering and model selection to construct a provably consistent algorithm (\textbf{WRMSM}) for the identification of high-dimensional fractal systems. In the threefold limit as dimension, sample size and scale go to infinity, the method consistently estimates the Hurst distribution. In addition, Monte Carlo simulations for realistic sample sizes demonstrate that the proposed methodology displays efficient finite-sample performance. The simulations also show that, in the context of the paper, \textbf{WRMSM} outperforms \textbf{GMM}, which is a popular clustering method. Use of the algorithm in the analysis of the large macroeconomic database called FRED-MD, held by the St.\ Louis Federal Reserve Bank, reveals evidence for the property of cointegration. 

The work contained in this paper leads to multiple new directions of both theoretical and applied research. On the theoretical side, one example of relevant challenge is the development of \textit{confidence sets} for the proposed algorithm, so as to quantify uncertainty in modeling practice. This, in turn, calls for new results on random matrix theory regarding local fluctuations of extreme wavelet log-eigenvalues (cf.\ Remark \ref{r:rates_of_convergence}, $(ii)$). Another challenge is related to modeling \textit{robustness} and \textit{universality}, i.e.,  the extension of the results to mathematical frameworks beyond mixed-Gaussian-type measurements of the form \eqref{e:Y(ell)=P(n)X(ell)_intro}. On the applied side, the tools developed in this paper are ripe for use in other promising areas of scientific research such as neuroscience, where the modeling of low-frequency brain connectivity is currently a cutting-edge research topic (cf.\ Ciuciu et al.\ \cite{ciuciu:abry:he:2014}, Qiu et al.\ \cite{qiu:han:liu:caffo:2016}, Hernandez et al.\ \cite{hernandez:etal:2023}, Dumeur \cite{dumeur:2025}).

\appendix

\section{Wavelet framework}

\subsection{Assumptions on the wavelet multiresolution analysis}\label{s:assumptions_on_the_MRA}

In the main results of the paper, we make use of the following conditions on the underlying wavelet multiresolution analysis (MRA).

\medskip

\noindent {\sc Assumption $(W1)$}: $\psi \in L^2(\bbR)$ is a wavelet function, namely, it satisfies the relations
\begin{equation}\label{e:N_psi}
\int_{\bbR} \psi^2(t)dt = 1 , \quad \int_{\bbR} t^{p}\psi(t)dt = 0, \quad p = 0,1,\hdots, N_{\psi}-1, \quad \int_{\bbR} t^{N_{\psi}}\psi(t)dt \neq 0,
\end{equation}
for some integer (number of vanishing moments) $N_{\psi} \geq 2$.

\medskip

\noindent {\sc Assumption ($W2$)}: the scaling and wavelet functions
\begin{equation}\label{e:supp_psi=compact}
\textnormal{$\phi\in L^1(\bbR)$ and $\psi\in L^1(\bbR)$ are compactly supported }
\end{equation}
and $\widehat{\phi}(0)=1$.

\medskip

\noindent {\sc Assumption $(W3)$}: there exist positive constants $A <\infty$ and $\alpha > 1$ such that
\begin{equation}\label{e:psihat_is_slower_than_a_power_function}
\sup_{x \in \bbR} |\widehat{\psi}(x)| (1 + |x|)^{\alpha} < A.
\end{equation}

%\medskip
%
%\noindent {\sc Assumption ($W4$)}: the function
%\begin{equation}\label{e:sum_k^m_phi(.-k)}
%\sum_{k\in \mathbb{Z}}k^m\phi(\cdot-k)
%\end{equation}
%is a polynomial of degree $m$ for all $m=0,\hdots,N_{\psi}-1$.

\medskip

Conditions \eqref{e:N_psi} and \eqref{e:supp_psi=compact} imply that $\widehat{\psi}(x)$ exists, is everywhere infinitely differentiable and its first $N_{\psi}-1$ derivatives are zero at $x = 0$. Condition \eqref{e:psihat_is_slower_than_a_power_function}, in turn, implies that $\psi$ is continuous (see Mallat \cite{mallat:1999}, Theorem 6.1) and, hence, bounded.

Note that assumptions ($W1-W3$) are closely related to the broad wavelet framework for the analysis of $\kappa$-th order ($\kappa \in \bbN \cup \{0\}$) stationary-increment stochastic processes laid out in Moulines et al.\ \cite{moulines:roueff:taqqu:2007:Fractals,moulines:roueff:taqqu:2007:JTSA,moulines:roueff:taqqu:2008} and Roueff and Taqqu~\cite{roueff:taqqu:2009}. The Daubechies scaling and wavelet functions generally satisfy ($W1-W3$) (see Moulines et al.\ \cite{moulines:roueff:taqqu:2008}, p.\ 1927, or Mallat \cite{mallat:1999}, p.\ 253). Usually, the parameter $\alpha$ increases to infinity as $N_{\psi}$ goes to infinity (see Moulines et al.\ \cite{moulines:roueff:taqqu:2008}, p.\ 1927, or Cohen \cite{cohen:2003}, Theorem 2.10.1). Also, under the orthogonality of the underlying wavelet and scaling function basis, ($W1-W3$) imply the so-called Strang-Fix condition (see Mallat \cite{mallat:1999}, Theorem 7.4, and Moulines et al.\ \cite{moulines:roueff:taqqu:2007:JTSA}, p.\ 159, condition (W-4)).

\subsection{Mallat's algorithm and discrete time measurements}\label{s:Mallats_algorithm}

Initially, suppose an infinite sequence of vector-valued measurements
\begin{equation}\label{e:infinite_sample_discrete}
\{Y(k)\}_{k \in \bbZ} \subseteq \bbR^p
\end{equation}
is available. Then, we can apply Mallat's algorithm to extract the so-named \textit{approximation} $({\mathcal A}(2^{j+1},\cdot))$ and \textit{detail} $({\mathcal D}(2^{j+1},\cdot))$ coefficients at coarser scales $2^{j+1}$ by means of an iterative procedure. In fact, as commonly done in the wavelet literature, we initialize the algorithm with the process
\begin{equation}\label{e:Btilde}
\bbR^p \ni \widetilde{Y}(t) := \sum_{k \in \bbZ} Y(k)\phi(t-k), \quad  t\in \bbR.
\end{equation}
By the orthogonality of the shifted scaling functions $\{\phi(\cdot - k)\}_{k \in \bbZ}$,
\begin{equation}\label{e:a(0,k)}
\bbR^p \ni  {\mathcal A}(2^0,k)= \int_\bbR \widetilde{Y}(t)\phi(t-k)dt= Y(k), \quad k \in \bbZ
\end{equation}
(see Stoev et al.\ \cite{stoev:pipiras:taqqu:2002}, proof of Lemma 6.1, or Moulines et al.\ \cite{moulines:roueff:taqqu:2007:JTSA}, p.\ 160; cf.\ Abry and Flandrin \cite{abry:flandrin:1994}, p.\ 33). In other words, the initial sequence, at octave $j = 0$, of approximation coefficients is given by the original sequence of random vectors. To obtain approximation and detail coefficients at coarser scales, we use Mallat's iterative procedure
\begin{equation}\label{e:Mallat}
{\mathcal A}(2^{j+1},k) = \sum_{k'\in \mathbb{Z}} u_{k'-2k}\hspace{0.5mm} {\mathcal A}(2^j,k'),\quad {\mathcal D}(2^{j+1},k) =\sum_{k'\in \mathbb{Z}}v_{k'-2k} \hspace{0.5mm} {\mathcal A}(2^j,k'), \quad j \in \mathbb{N} \cup \{0\}, \quad k \in \mathbb{Z},
\end{equation}
where the (scalar) filter sequences $\{ u_k:=2^{-1/2}\int_\bbR \phi(t/2)\phi(t-k)dt \}_{k\in\bbZ}$, $\{v_k:=2^{-1/2}\int_\bbR\psi(t/2)\phi(t-k)dt\}_{k\in\bbZ}$ are called low- and high-pass MRA filters, respectively. Due to the assumed compactness of the supports of $\psi$ and of the associated scaling function $\phi$ (see condition \eqref{e:supp_psi=compact}), only a finite number of filter terms is nonzero, which is convenient for computational purposes (Daubechies \cite{daubechies:1992}). So, assume without loss of generality that $\text{supp}(\phi) = \text{supp}(\psi)=[0,T]$ (cf.\ Moulines et al.\ \cite{moulines:roueff:taqqu:2007:JTSA}, p.\ 160). Then, the wavelet (detail) coefficients ${\mathcal D}(2^j,k)$ of $Y$ can be expressed as \eqref{e:disc2}, where the filter terms are defined by \eqref{e:hj,l}.

If we replace \eqref{e:infinite_sample_discrete} with the realistic assumption that only a finite length series \eqref{e:finite_sample} is available, writing $\widetilde{Y}^{(n)}(t) := \sum_{k=1}^{n}Y(k)\phi(t-k)$, we have $\widetilde{Y}^{(n)}(t) = \widetilde{Y}(t)$ for all $t\in (T,n+1)$ (cf.\ Moulines et al.\ \cite{moulines:roueff:taqqu:2007:JTSA}). Noting that ${\mathcal D}(2^j,k) = \int_\bbR \widetilde Y(t) 2^{-j/2}\psi(2^{-j}t - k)dt$ and ${\mathcal D}^{(n)}(2^j,k) = \int_\bbR \widetilde Y^{(n)}(t)2^{-j/2}\psi(2^{-j}t - k)dt$, it follows that the finite-sample wavelet coefficients ${\mathcal D}^{(n)}(2^j,k)$ of $\widetilde{Y}^{(n)}(t) $ are equal to ${\mathcal D}(2^j,k)$ whenever $\textnormal{supp } \psi(2^{-j} \cdot - k) = (2^jk, 2^j (k+T)) \subseteq (T,n+1)$. In other words,
\begin{equation}\label{e:d^n=d}
{\mathcal D}^{(n)}(2^j,k) = {\mathcal D}(2^j,k),\qquad \forall (j,k)\in\{(j,k): 2^{-j}T\leq k\leq 2^{-j}(n+1)-T\}.
\end{equation}
Equivalently, such subset of finite-sample wavelet coefficients is not affected by the so-named \textit{border effect} (cf.\ Craigmile et al.\ \cite{craigmile:guttorp:Percival:2005}, Percival and Walden \cite{percival:walden:2006}). Moreover, by \eqref{e:d^n=d} the number of such coefficients at octave $j$ is given by $n_j = \lfloor 2^{-j}(n+1-T)-T\rfloor$. Hence, $n_j \sim 2^{-j}n$ for large $n$. Thus, for notational simplicity we suppose
 \begin{equation}\label{e:nj=n/2^j}
 n_{j} = \frac{n}{2^j} \in \bbN
 \end{equation}
 holds exactly and only work with wavelet coefficients unaffected by the border effect.

\section{Main proofs and auxiliary results}

In this section, we provide the proof of the main result of the paper (Theorem \ref{t:WRMSM_consistency}), and also state and prove some key auxiliary results. Recall that, whenever convenient, we may write $a = a(n)$ and $p = p(n)$ for notational simplicity (cf.\ \eqref{e:a=a(n)_or_p=p(n)}). 

In the following proposition, we establish that, as long as one starts with a ``good" choice of hyperparameter $\varepsilon$, \textbf{HD$\varepsilon$ES} is, itself, a consistent method for the estimation of the Hurst distribution. Most importantly, Proposition \ref{t:fixed_epsilon_consistency} is used directly in the proof of Theorem \ref{t:WRMSM_consistency}.
\begin{proposition}\label{t:fixed_epsilon_consistency}
Suppose assumptions ($A1-A5$) and ($W1-W3$) hold. Let $\widehat{\mathsf{H}}_p$ be as in \eqref{e:scaling_estimate} and let $\Delta_{\min}$ be as in \eqref{e:threshold_0}. Fix 
\begin{equation}\label{e:varepsilon_in_(0,Delta-min)}
\varepsilon \in (0, \Delta_{\min}).  
\end{equation}
Let $\widehat{r}_\varepsilon$, $\{ \widehat{\breve{H}}_1,\hdots,\widehat{\breve{H}}_{\widehat{r}_{\varepsilon}}\}$, and $\{ \widehat{\pi}_1,\hdots,\widehat{\pi}_{\widehat{r}_{\varepsilon}}\}$ be estimators obtained by means of \textbf{HD$\varepsilon$ES}. Then, 
    for any $\eta >0$, as $n\to\infty$,
    \begin{equation}\label{e:good_epsilon_consistency}
    \bbP\Big( \{\widehat{r}_{\varepsilon} = r\} \bigcap \Big\{  \max_{i = 1,\hdots, r}| \widehat{\breve{H}}_i -\breve{H}_i| < \eta \Big\}  \bigcap \Big\{ \max_{i = 1,\hdots, r}| \widehat{\pi}_i -\pi(\breve{H}_i)| < \eta \Big\} \Big) \to 1.
    \end{equation}
\end{proposition}

%For theoretical guarantees, we require that $M$ and $m$ be chosen in such a way that the refined parameter search space contain at least one candidate threshold parameter less than $\Delta_{\min}$. The following theorem demonstrates that \textbf{Step 2} of the proposed algorithm is consistent in obtaining a ``good" choice of $\varepsilon \in (0,\Delta_{\min})$. Not less importantly, it further demonstrates that the overall WRMSM algorithm (i.e., \textbf{Steps 1--2}) is consistent (on excluding the value $\varepsilon = \Delta_{\min}$ in \eqref{e:E=choice_set}, see Remark \ref{r:varepsilon=Delta}).
%
%\begin{proposition}\label{t:model_selection_is_consistent} Suppose the assumptions of Theorem \ref{t:WRMSM_consistency} hold. Then,
%    \begin{equation}\label{e:good_model_selected_threshold}
%    \bbP(\varepsilon_{ms} < \Delta_{\min})\to 1,\quad n\to\infty.
%    \end{equation}
%In addition, \eqref{e:good_epsilon_consistency_ms} also holds. 
%\end{proposition}
%
%
%
%Relation \eqref{e:good_epsilon_consistency_ms}  clearly mirrors relation \eqref{e:good_epsilon_consistency}, except that model selection (choice of $\varepsilon > 0$) is also contemplated in the former.\\

In order to show Proposition \ref{t:fixed_epsilon_consistency}, we will need the following two lemmas, i.e., Lemmas \ref{p:log_eigenvalues_count} and \ref{p:ball_average_and_proportion}.

In the first one, we make use of results from Abry et al.\ \cite{abry:didier:orejola:wendt:2024}. The lemma establishes that, for any ``good" threshold, $\varepsilon\in(0,\Delta_{\min})$, with increasing probability the wavelet log-eigenvalues \eqref{e:scaling_estimate} eventually coalesce within balls of radius $\varepsilon$ around each point in the support of the Hurst exponent distribution. As a consequence, the intersection of $\widehat{\mathsf{H}}_{p}$ with those balls eventually forms a \textit{partition} of  $\widehat{\mathsf{H}}_{p}$. In order to state the lemma, let $\mathsf{H}_p$ be the diagonal entries of  the scaling matrix $\mathds{H}_n$ as in \eqref{e:mathds_hn}. Namely,
\begin{equation}\label{e:sampled_exponenets}
    \mathsf{H}_p:=\{ H_1,\hdots,H_{p(n)}\}.
\end{equation}

\begin{lemma}\label{p:log_eigenvalues_count}
Suppose conditions ($A1-A5$) and ($W1-W3$) hold. Let $\widehat{\mathsf{H}}_p$ be as in \eqref{e:scaling_estimate} and  $\mathsf{H}_p$ be as in \eqref{e:sampled_exponenets}. Also, consider $\varepsilon$ satisfying \eqref{e:varepsilon_in_(0,Delta-min)}. Then, with probability going to 1, 
\begin{equation}\label{e:card_rescaled_log-eig=card_H=Hbreve}
|\widehat{\mathsf{H}}_{p} \cap {\mathcal B}_{\varepsilon}(\breve{H}_i)|= \sum_{H\in\mathsf{H}_p}\mathds{1}_{\{H =\breve{H}_i\}}, \quad i= 1,\hdots,r.
\end{equation}
 In addition, for each pair $1\leq i < j\leq r$,
\begin{equation}\label{e:intersection_converges_to_zero}
\big|\widehat{\mathsf{H}}_{p} \cap {\mathcal B}_{\varepsilon}(\breve{H}_i)\bigcap\widehat{\mathsf{H}}_{p} \cap {\mathcal B}_{\varepsilon}(\breve{H}_j)\big| = 0
\end{equation}
with probability tending towards 1.
\end{lemma}
\begin{proof}
First consider the case where 
\begin{equation}\label{e:varepsilon_in_(0,Delta/2)}
\varepsilon \in (0,\Delta_{\min}/2). 
\end{equation}
By expression (3.32) in the proof of Theorem 3.3 in Abry et al.\ \cite{abry:didier:orejola:wendt:2024} (see Theorem \ref{t:main_theorem_discrete} in this paper), for every $n$ and each $\ell = 1,\hdots, p(n)$,
\begin{equation}
\label{e:log_eigenvalue_bound1}
   G_{L}(n)+H_{(\ell)} \leq \widehat{H}_\ell
\leq  H_{(\ell)} + G_{U}(n) \quad \textnormal{a.s.}
\end{equation}
In \eqref{e:log_eigenvalue_bound1}, we define
\begin{equation}\label{e:G_lower}
 G_{L}(n) := \frac{\log \lambda_{1} \big( \boldsymbol{\mathcal{W}}_X(2^j)\big)  }{2\log a(n)}
\end{equation}
and
\begin{equation} \label{e:G_upper}
       G_{U}(n) :=  \frac{\log \lambda_{p} \big( \boldsymbol{\mathcal{W}}_X(2^j)\big)  }{2\log a(n)}.
\end{equation}
It is further shown in Abry et al.\ \cite{abry:didier:orejola:wendt:2024}  that
\begin{equation}
\label{e:G_limit}
\max\{ | G_{L}(n) |, |G_{U}(n)| \} \stackrel{\bbP}\to 0, \quad n \rightarrow \infty.
\end{equation}
So, let 
\begin{equation}\label{e:A_n,varepsilon}
A_{n,\varepsilon} = \big\{\omega: \max\{|G_{L}(n)|,|G_{U}(n)|\} \leq \varepsilon \big\}. 
\end{equation}
Then, for any fixed $\delta > 0$, relation \eqref{e:G_limit} implies that, for large enough $n$, $\bbP\big( A_{n,\varepsilon}\big) > 1- \delta$. Therefore, for any $\omega \in A_{n,\varepsilon}$, we can recast \eqref{e:log_eigenvalue_bound1} in the form
\begin{equation} \label{e:log_eigenvalue_bound2}
|H_{(\ell)} - \widehat{H}_\ell| \leq  \varepsilon, \quad \ell=1,\hdots,p.
\end{equation}
Moreover, under \eqref{e:varepsilon_in_(0,Delta/2)}, 
\begin{equation}\label{e:disjoint_balls}
\mathcal{B}_\varepsilon(\breve{H}_i) \cap \mathcal{B}_\varepsilon(\breve{H}_j ) = \emptyset, \quad i \neq j. 
\end{equation}
As a consequence of \eqref{e:log_eigenvalue_bound2} and \eqref{e:disjoint_balls}, both relations \eqref{e:card_rescaled_log-eig=card_H=Hbreve} and \eqref{e:intersection_converges_to_zero} hold under \eqref{e:varepsilon_in_(0,Delta/2)}. 
%$$
%|\widehat{\mathsf{H}}_p\cap \mathcal{B}_\varepsilon(\breve{H}_i)| = \sum_{H\in\mathsf{H}_p}\mathds{1}_{\{H =\breve{H}_i\}}.
%$$
%Hence, with probability going to 1, \OO{expression \eqref{e:card_rescaled_log-eig=card_H=Hbreve} holds for $\varepsilon\in(0,\Delta_{\min}/2)$.} 

%Additionally, observe that
%\begin{equation}\label{e:p_total_estimtes}
% \sum_{i=1}^r \sum_{H\in\mathsf{H}_p}\mathds{1}_{\{H =\breve{H}_i\}} = p = p(n) \quad \textnormal{a.s.}, \quad n \in \bbN, 
%\end{equation}
%where the equality follows from the fact that the double summation counts the number of modes of each type. Therefore, as a consequence of \eqref{e:card_rescaled_log-eig=card_H=Hbreve} and \eqref{e:p_total_estimtes}, \OO{
% \begin{equation}\label{e:delta/withprob1}
% \bbP\big(\{\omega \ : \{ \widehat{\mathsf{H}}_p\cap\mathcal{B}_\varepsilon(\breve{H}_1),\hdots,\widehat{\mathsf{H}}_p\cap\mathcal{B}_\varepsilon(\breve{H}_r)\} \textnormal{ forms a partition of } \widehat{\mathsf{H}}_p \}\big)\to 1,\quad n\to\infty. \end{equation}
%Hence \eqref{e:intersection_converges_to_zero} holds for $\varepsilon\in(0,\Delta_{\min}/2).$}

Alternatively, consider the case where 
\begin{equation}\label{e:varepsilon>=Deltamin/2}
\varepsilon \in [\Delta_{\min}/2,\Delta_{\min}). 
\end{equation}
So, let  $\varepsilon' \in (0,\Delta_{\min}/2)$. By the previous case \eqref{e:varepsilon_in_(0,Delta/2)}, with probability converging to 1 relations \eqref{e:log_eigenvalue_bound2} and \eqref{e:disjoint_balls} hold with $\varepsilon'$ in place of $\varepsilon$. Hence, for each $i=1,...,r$, and for any $j\neq i$, 
$$\widehat{\mathsf{H}}_p\cap\mathcal{B}_{\varepsilon'}(\breve{H}_i) \subseteq \widehat{\mathsf{H}}_p\cap\mathcal{B}_{\varepsilon}(\breve{H}_i) \quad \textnormal{and} \quad |\widehat{\mathsf{H}}_p\cap\mathcal{B}_{\varepsilon'}(\breve{H}_i) \bigcap\widehat{\mathsf{H}}_p\cap\mathcal{B}_{\varepsilon}(\breve{H}_j)|=0.
$$
Consequently, relations \eqref{e:card_rescaled_log-eig=card_H=Hbreve} and \eqref{e:intersection_converges_to_zero} also hold, with probability tending to 1, under \eqref{e:varepsilon>=Deltamin/2}. This establishes the proposition. $\Box$ \\
%Therefore, if the set $\{ \widehat{\mathsf{H}}_p\cap\mathcal{B}_{\varepsilon'}(\breve{H}_1),\hdots,\widehat{\mathsf{H}}_p\cap\mathcal{B}_{\varepsilon'}(\breve{H}_r)\}$ forms a partition of $ \widehat{\mathsf{H}}_p$, then the set $\{ \widehat{\mathsf{H}}_p\cap\mathcal{B}_{\varepsilon}(\breve{H}_1),\hdots,\widehat{\mathsf{H}}_p\cap\mathcal{B}_{\varepsilon}(\breve{H}_r)\}$ does as well. Since $\varepsilon' \in (0,\Delta_{\min}/2)$,  \eqref{e:card_rescaled_log-eig=card_H=Hbreve} and \eqref{e:intersection_converges_to_zero} hold for $\varepsilon\in[\Delta_{\min}/2,\Delta_{\min})$. 
%
\end{proof}

The next lemma is also used in the proof of Proposition \ref{t:fixed_epsilon_consistency}. It shows that, for any Hurst mode $\breve{H}_i$, $i=1,\hdots,r$, the proportion of wavelet log-eigenvalues $\widehat{H}_{\ell}$, $\ell = 1,\hdots,p$, within any ball with small enough radius centered around $\breve{H}_i$ converges in probability to $\pi(\breve{H}_i)$. It further shows that the average value of the wavelet log-eigenvalues in each of these balls converges to a Hurst mode.  
\begin{lemma}\label{p:ball_average_and_proportion}
Suppose conditions ($A1-A5$) and ($W1-W3$) hold. Let $\widehat{\mathsf{H}}_p $ be as in \eqref{e:scaling_estimate} and let $\mathsf{H}_p$ be as in \eqref{e:sampled_exponenets}. Also, consider $\varepsilon$ satisfying \eqref{e:varepsilon_in_(0,Delta-min)}.  In addition, let
\begin{equation}\label{e:ball_converges_to_proportion}
\widehat{p}_i = \widehat{p}_i(n):= |\widehat{\mathsf{H}}_p\cap \mathcal{B}_\varepsilon(\breve{H}_i)| \in \bbN \cup \{0\}, \quad i=1,\hdots,r.
\end{equation}
Then, with probability going to 1 as $n \rightarrow \infty$, for any $i= 1,\hdots,r$,
\begin{equation}\label{e:proportion_ball_converges_to_probability}
    \frac{\widehat{p}_i(n) }{p(n)} \stackrel{\bbP}\to \pi(\breve{H}_i)
\end{equation}
and 
\begin{equation}\label{e:mean_of_ball_converges}
\textnormal{mean} (\widehat{\mathsf{H}}_{p} \cap {\mathcal B}_{\varepsilon}(\breve{H}_i)) \stackrel{\bbP}\to \breve{H}_i.
\end{equation}
\end{lemma}
\begin{proof}
We first show \eqref{e:proportion_ball_converges_to_probability}. Recall that the entries of the vector \eqref{e:sampled_exponenets} are obtained independently (see assumption ($A2$) and expression \eqref{e:mathds_hn}). Then, by Kolmogorov's strong law of large numbers,
\begin{equation}\label{e:propriton_converges_LLN}
\frac{1}{p}\sum_{H\in\mathsf{H}_p}\mathds{1}_{\{H =\breve{H}_i\}}  \stackrel{a.s.}\to \pi(\breve{H}_i),\quad n\to\infty.
\end{equation}
Relation \eqref{e:proportion_ball_converges_to_probability} is now a consequence of Lemma \ref{p:log_eigenvalues_count} (see expression \eqref{e:card_rescaled_log-eig=card_H=Hbreve}).  

Turning to \eqref{e:mean_of_ball_converges}, without loss of generality consider any $\delta \in \big(0, \min\{ \varepsilon ,\Delta_{\min}/2\} \big)$ (cf.\ \eqref{e:varepsilon_in_(0,Delta/2)}). Then, as in the proof of Lemma \ref{p:log_eigenvalues_count} (see expression \eqref{e:log_eigenvalue_bound2}), we have that, with probability converging to 1,
$$
\breve{H}_i - \delta \leq \textnormal{mean}(\widehat{\mathsf{H}}_{p} \cap {\mathcal B}_{\delta}(\breve{H}_i)) \leq  \breve{H}_i + \delta.
$$
However, since $\mathcal{B}_\delta(\breve{H}_i)\subseteq \mathcal{B}_\varepsilon(\breve{H}_i)$, it follows that $\widehat{\mathsf{H}}_{p} \cap {\mathcal B}_{\varepsilon}(\breve{H}_i) = \widehat{\mathsf{H}}_{p} \cap {\mathcal B}_{\delta}(\breve{H}_i)$ with probability approaching 1. Hence,
$$
\bbP\big(| \textnormal{mean}(\widehat{\mathsf{H}}_{p} \cap {\mathcal B}_{\varepsilon}(\breve{H}_i)) -  \breve{H}_i | \leq \delta  \big)\to 1, \quad n \rightarrow \infty,
$$
establishing \eqref{e:mean_of_ball_converges}. This completes the proof. $\Box$
\end{proof}

\medskip
We are now in a position to prove Proposition \ref{t:fixed_epsilon_consistency}.

\medskip
\noindent{\sc Proof of Proposition \ref{t:fixed_epsilon_consistency}}:  Let $\widehat{r}_{\varepsilon}$ be as defined in \textbf{HD$\varepsilon$ES}. We first show that, with probability going to $1$,
    \begin{equation}\label{e:r_hat_estimate_is_consistent}
        \widehat{r}_{\varepsilon}= r.
    \end{equation}
    For $\varepsilon$ as in \eqref{e:varepsilon_in_(0,Delta-min)}, consider the graph induced by an $\varepsilon$-threshold $G_\varepsilon(\widehat{\mathsf{H}}_p)$. Lemma \ref{p:log_eigenvalues_count} implies that, with probability tending to $1$, we can write the decomposition
    \begin{equation}\label{e:G_decomp_disjoint_complete_graphs}
    G_\varepsilon(\widehat{\mathsf{H}}_p) = \bigcupdot^{r}_{i=1} G_i.
    \end{equation}
In \eqref{e:G_decomp_disjoint_complete_graphs}, $G_i = (E_i,V_i)$, $i = 1,\hdots,r$, are complete and pairwise disjoint (sub)graphs.  For each $i$, we define that the vertices in $G_i$ correspond to the points in $ \widehat{\mathsf{H}}_p\cap \mathcal{B}_\varepsilon(\breve{H}_i)$, i.e., we write 
\begin{equation}\label{e:graph_decomp_indexing}
\widehat{H}_\ell \in \widehat{\mathsf{H}}_p\cap \mathcal{B}_\varepsilon(\breve{H}_i) \Leftrightarrow \ell \in V_i.
 \end{equation}
 Also, let ${\mathbf L}_\varepsilon(\widehat{\mathsf{H}}_p)$ be the graph Laplacian as in \eqref{e:graph_laplacian_mat}, and recall that
\begin{equation}\label{e:graph_Laplacian_eigenvalues}
\theta_1 = \theta_1\big({\mathbf L}_\varepsilon(\widehat{\mathsf{H}}_p) \big) \leq \hdots \leq \theta_p = \theta_p \big({\mathbf L}_\varepsilon(\widehat{\mathsf{H}}_p)\big)
\end{equation}
denote its ordered eigenvalues (see \textbf{HD$\varepsilon$SE}). As a consequence of \eqref{e:G_decomp_disjoint_complete_graphs} and Lemma \ref{l:null_space_of_graph_laplacian}, ${\mathbf L}_\varepsilon(\widehat{\mathsf{H}}_p)$ has a zero eigenvalue with multiplicity $r$ with probability going to 1 as $n \rightarrow \infty$. Moreover, the non-zero eigenvalues of ${\mathbf L}_\varepsilon(\widehat{\mathsf{H}}_p)$ take values 
\begin{equation}\label{e:p-hat_ell_values}
\widehat{p}_1,\hdots,\widehat{p}_r
\end{equation}
(see \eqref{e:ball_converges_to_proportion}), each with multiplicity 
\begin{equation}\label{e:p-hat_ell-1_multiplicity}
\widehat{p}_1-1,\hdots,\widehat{p}_r-1, 
\end{equation}
respectively. So, let $\widehat{p}_{(1)} \leq \hdots \leq \widehat{p}_{(r)}$ be the associated ordered $r$-tuple. We obtain
\begin{equation}\label{e:eigen_gap_converges}
\bbP\big(\theta_{r+1} -\theta_{r}  = \widehat{p}_{(1)}\big)\to 1, \quad n \rightarrow \infty.
\end{equation}
On the other hand, by Lemma \ref{p:ball_average_and_proportion} (see \eqref{e:proportion_ball_converges_to_probability}) and assumption ($A1$) (see expression \eqref{e:difference_of_probabilites_bounded}),
\begin{equation}\label{e:proportion_gap}
\bbP\Big( \widehat{p}_{(1)} \geq \max_{i = 1,\hdots,r-1}\{\widehat{p}_{(i+1)} -\widehat{p}_{(i)}\}\Big) \to 1,\quad n\to\infty.
\end{equation}
Therefore, from expressions \eqref{e:eigen_gap_converges} and \eqref{e:proportion_gap},
\begin{equation}\label{e:P(theta_r+1-theta_r>= max)}
\bbP\Big( \theta_{r+1} -\theta_{r} \geq \max_{i = 1,\hdots, r-1}\{\widehat{p}_{(i+1)} -\widehat{p}_{(i)}\}\Big) \to 1,\quad n\to\infty.
\end{equation}
As a consequence of \eqref{e:p-hat_ell_values}, \eqref{e:p-hat_ell-1_multiplicity} and \eqref{e:P(theta_r+1-theta_r>= max)}, $\widehat{r}_\varepsilon := \textnormal{argmax}_{i = 1,\hdots,  p(n)-1} \big\{ \theta_{i+1}  -  \theta_{i} \big\} \stackrel{\bbP}\to r$ as $n\to\infty$ (\textbf{n.b.}: $\textnormal{argmax} \neq \max$).  This establishes \eqref{e:r_hat_estimate_is_consistent}.\\

Next, we show that, for any $\eta >0$, as $n\to\infty$,
    \begin{equation}\label{e:good_epsilon_consistency_ms_proof}
    \bbP\Big( \Big\{ \max_{i = 1,\hdots, r}| \widehat{\breve{H}}_i -\breve{H}_i| < \eta \Big\}  \bigcap \Big\{ \max_{i = 1,\hdots, r}| \widehat{\pi}_i -\pi(\breve{H}_i)| < \eta \Big\} \Big) \to 1.
    \end{equation}
Consider a sequence of events
\begin{equation}\label{e:good_set_r_disjoint}
\widetilde{A}_{n,\varepsilon} := \big\{
\textnormal{$\omega$ : \textnormal{following \textbf{WRMSM}, } $\widehat{r}_\varepsilon = r$ and \eqref{e:G_decomp_disjoint_complete_graphs} holds} \big\}.
\end{equation}
Note that 
\begin{equation}\label{e:P(A_n,varepsilon)->1}
\bbP(\widetilde{A}_{n,\varepsilon})\to1,  \quad n\to\infty. 
\end{equation}
In consideration of any 
\begin{equation}\label{e:omega_in_An,epsilon}
\omega \in \widetilde{A}_{n,\varepsilon}, 
\end{equation}
let $\mathbf{u}_1,\hdots,\mathbf{u}_r \in \bbR^p$ be the eigenvectors corresponding to the zero eigenvalue of ${\mathbf L}_\varepsilon(\widehat{\mathsf{H}}_p)$. Also, define the matrix
$$
\bbR^{p \times r} \ni {\mathbf U} = \big(\mathbf{u}_1,\hdots,\mathbf{u}_r\big) 
$$
and let $\mathbf{y}_i$, $i = 1,\hdots,p$, be the rows of ${\mathbf U}$. Then, still under \eqref{e:omega_in_An,epsilon}, Lemma \ref{l:null_space_of_graph_laplacian} implies that the set
$$
S := \{\mathbf{y}_1,\hdots,\mathbf{y}_p\}
$$
consists of exactly $r$ \textit{distinct} vectors in $\mathbb{R}^{r}$. Bearing in mind relation \eqref{e:G_decomp_disjoint_complete_graphs}, this is so because, for each $i = 1,\hdots,r$, $\ell,\ell' \in V_i$ if and only if $\mathbf{y}_{\ell}=\mathbf{y}_{\ell'}$. Therefore, from Lemma \ref{l:k-means_converges}, it follows that an application of the $k$-means algorithm to the set $S$ and for the choice $\kappa :=\widehat{r}_\varepsilon$ yields the clusters
$$
C'_1 = \{ \mathbf{y}_\ell  : \ell \in V_1\},\hdots,C'_{\widehat{r}_\varepsilon} = \{ \mathbf{y}_\ell : \ell \in V_r\}.
$$
Hence, by \eqref{e:G_decomp_disjoint_complete_graphs}, under \eqref{e:omega_in_An,epsilon} the clusters obtained in \textbf{HD$\varepsilon$SE} are given by
\begin{equation}\label{e:good_clusters}
    \mathcal{C}_\varepsilon = \Big\{ C_i = \widehat{\mathsf{H}}_p\cap \mathcal{B}_\varepsilon(\breve{H}_i), \hspace{1mm}i=1,\hdots,r, \textnormal{ where }\widehat{\mathsf{H}}_p = \bigcupdot^{r}_{i=1} C_i\Big\}.
\end{equation}
In turn, still under \eqref{e:omega_in_An,epsilon}, relation \eqref{e:good_clusters} implies that the estimators 
$\widehat{\breve{H}}_i := \textnormal{mean}(C_i)$ for $\breve{H}_i$ satisfy
\begin{equation}\label{e:mean_is_in_a_ball}
\widehat{\breve{H}}_i =\textnormal{mean}\big(\widehat{\mathsf{H}}_p \cap \mathcal{B}_\varepsilon(\breve{H}_i)\big), \quad i = 1,\hdots,r = \widehat{r}_\varepsilon.
\end{equation}
Relation \eqref{e:good_clusters} further entails that the estimators $\widehat{\pi}_i := |C_i|/{p}$ for $\pi(\breve{H}_i)$ (see \textbf{HD$\varepsilon$SE}) satisfy
\begin{equation}\label{e:proportion_est_is_ball}
\widehat{\pi}_i = \frac{|\widehat{\mathsf{H}}_{p} \cap {\mathcal B}_{\varepsilon}(\breve{H}_i)|}{p}, \quad i = 1,\hdots, r.
\end{equation}
However, by Lemma \ref{p:ball_average_and_proportion}, for any $\eta>0$, as $n\to\infty$,
\begin{equation}\label{e:P(max)->1}
\bbP\Big(\Big\{ \max_{i = 1,\hdots, r} \Big| \frac{|\widehat{\mathsf{H}}_{p} \cap {\mathcal B}_{\varepsilon}(\breve{H}_i)|}{p} - \pi(\breve{H}_i)\Big| < \eta\Big\} \bigcap \Big\{ \max_{i = 1,\hdots, r} \Big| \textnormal{mean}\big(\widehat{\mathsf{H}}_p \cap \mathcal{B}_\varepsilon(\breve{H}_i)\big) -\breve{H}_i\Big| < \eta\Big\}  \Big) \to 1 .
\end{equation}
Therefore, in consideration of \eqref{e:P(A_n,varepsilon)->1}, \eqref{e:mean_is_in_a_ball}, \eqref{e:proportion_est_is_ball} and \eqref{e:P(max)->1}, it follows that \eqref{e:good_epsilon_consistency_ms_proof} holds.

So, as a consequence of \eqref{e:r_hat_estimate_is_consistent} and \eqref{e:good_epsilon_consistency_ms_proof}, \eqref{e:good_epsilon_consistency} holds. This completes the proof. $\Box$\\

The next lemma demonstrates that $\textnormal{ICSD}$ for clusters obtained by ``good" choices of $\varepsilon$ vanish. Conversely, it also shows that it does not converge for ``bad" choices of $\varepsilon$. The lemma is used to establish Theorem \ref{t:WRMSM_consistency}. The proof makes use of constructions put forth in the proof of Lemma \ref{p:ball_average_and_proportion} (on the choice $\varepsilon = \Delta_{\min}$, see Remark \ref{r:varepsilon=Delta}).
\begin{lemma}\label{p:ICDS_to_zero}
Suppose assumptions ($A1-A5$) and ($W1-W3$) hold. Let $\widehat{\mathsf{H}}_p$ be as in \eqref{e:scaling_estimate} and let $\Delta_{\min}$ be as in \eqref{e:threshold_0}.
\begin{enumerate}
    \item[$(i)$] Fix 
    \begin{equation}\label{e:varepsilon_in_(0,Delta_min)}
    \varepsilon_g \in (0, \Delta_{\min}). 
    \end{equation}
     Let $\mathcal{C}_{\varepsilon_g}$ be the clustering obtained by \textbf{HD$\varepsilon$SE}. Then,
    \begin{equation}\label{e:good_ICSD_to_zero}
    \textnormal{ICSD}(\mathcal{C}_{\varepsilon_g})\stackrel{\mathbb{P}}\to 0,\quad n\to\infty.
    \end{equation}
    \item[$(ii)$] Fix 
    \begin{equation}\label{e:varepsilonb>Delta_min}
    \varepsilon_{b} > \Delta_{\min}. 
    \end{equation}
    Let $\mathcal{C}_{\varepsilon_b}$ be the clustering obtained by \textbf{HD$\varepsilon$SE}. Then, there exists $\Upsilon > 0$ such that
    \begin{equation}\label{e:bad_ICSD_not_to_zero}
    \textnormal{ICSD}(\mathcal{C}_{\varepsilon_b})\geq \Upsilon + o_{\bbP}(1).
    \end{equation}
\end{enumerate}
\end{lemma}
\begin{proof}
    We first show ($i$). Under \eqref{e:varepsilon_in_(0,Delta_min)}, it suffices to show that, for any arbitrarily small
     \begin{equation}\label{e:varepsilon_in_(0,varepsilon)}
    0 < \delta < \min \big\{ \Delta_{\min} , \Delta_{\min} - \varepsilon_g \big\},
    \end{equation}
    the bound
    \begin{equation}\label{e:ICSD(Cdelta)<small}
    \textnormal{ICSD}(\mathcal{C}_{\varepsilon_g})  \leq r \cdot \delta + o_{\bbP}(1)
    \end{equation}
    holds with probability tending towards 1. 
    
    So, fix $\delta$ as in \eqref{e:varepsilon_in_(0,varepsilon)} and let $\widetilde{A}_{n,\delta}$ be as in \eqref{e:good_set_r_disjoint} (after replacing $\varepsilon$ with $\delta$). For $\omega \in \widetilde{A}_{n,\delta}$, the clusters obtained in \textbf{HD$\varepsilon$SE} are given by ${\mathcal C}_{\delta}$ as in \eqref{e:good_clusters} (again, after replacing $\varepsilon$ with $\delta$). In view of \eqref{e:varepsilon_in_(0,varepsilon)}, for $\omega \in \widetilde{A}_{n,\delta}$ the clusters obtained in \textbf{HD$\varepsilon$SE} based on $\varepsilon_g$ as in \eqref{e:varepsilon_in_(0,Delta_min)} are given by ${\mathcal C}_{\varepsilon_g} =  {\mathcal C}_{\delta}$. In particular, 
    \begin{equation}\label{e:ICSD(C_varepsilon)=ICSD(C_delta)}
    \textnormal{ICSD}(\mathcal{C}_{\varepsilon_g}) = \textnormal{ICSD}(\mathcal{C}_{\delta}).
    \end{equation}
    %    Let $\widetilde{A}_{n,\delta}$ and $\widetilde{A}_{n,\varepsilon}$ be events as in \eqref{e:good_set_r_disjoint}. Recall that, for $\omega$ in each of these two events, we obtain clusters $\mathcal{C}_\delta$ and $\mathcal{C}_\varepsilon$, respectively, as in \eqref{e:good_clusters}.  Now, note that, since $\mathcal{B}_\delta(\breve{H}_i)\subseteq \mathcal{B}_\varepsilon(\breve{H}_i)$, $i=1,\hdots,r$,  then $\widetilde{A}_{n,\delta}\subseteq \widetilde{A}_{n,\varepsilon}$. Therefore, for any $\omega \in \widetilde{A}_{n,\delta}$,
%\begin{equation}\label{e:two_balls_same_ICDS}
%\textnormal{ICSD}(\mathcal{C}_\varepsilon) = \textnormal{ICSD}(\mathcal{C}_\delta).
%\end{equation}
So, let $A_{n,\delta}$ be the event as in \eqref{e:A_n,varepsilon} (after replacing $\varepsilon$ with $\delta$). Still for $\delta$ as in \eqref{e:varepsilon_in_(0,varepsilon)}, define the events
$$
A^*_{n,\delta} = A_{n,\delta} \cap\widetilde{A}_{n,\delta}, \quad n \in \bbN.
$$
Then, by \eqref{e:G_limit} and \eqref{e:omega_in_An,epsilon}, $\bbP(A^*_{n,\delta})\rightarrow 1$ as $n \rightarrow \infty$. For any $n \in \bbN$ and $\omega \in A^*_{n,\delta}$, consider the associated clusters \eqref{e:good_clusters}. For notational simplicity, recast $ \mathcal{C}_\delta = \big\{ C_{\delta,i}, i =1\hdots,r \big\}$. Note that, again for any $\omega \in A^*_{n,\delta}$, $|H_{(\ell)}- \widehat{H}_{\ell}| \leq \delta$, $\ell = 1,\hdots, p$. Therefore, for any $i = 1,\hdots, r$, 
\begin{equation}\label{e:H-hat_in_C_<=>_H_(ell)_in_H-breve}
\widehat{H}_{\ell} \in C_{\delta,i} \hspace{1mm} \Leftrightarrow  \hspace{1mm} H_{(\ell)} = \breve{H}_i. 
\end{equation}
Hence,
\begin{equation}\label{e:sum_ell:H-ell_in_Cdelta,i=0|}
\sum_{\ell: \hspace{0.5mm}\widehat{H}_\ell \in C_{\delta,i}}|\breve{H}_i - H_{(\ell)}  |^2  = 0, \quad i = 1,\hdots, r.
\end{equation}
For any $i = 1,\hdots,r$, by Minkowski's inequality, relations \eqref{e:sum_ell:H-ell_in_Cdelta,i=0|} and \eqref{e:H-hat_in_C_<=>_H_(ell)_in_H-breve}, and by Lemma \ref{p:ball_average_and_proportion} (see \eqref{e:mean_of_ball_converges}),
$$
\Bigg( \sum_{\ell: \hspace{0.5mm}\widehat{H}_\ell \in C_{\delta,i}}|\widehat{H}_\ell - \textnormal{mean}(C_{\delta,i})|^2 \Bigg)^{1/2}
\leq \Bigg( \sum_{\ell: \hspace{0.5mm}\widehat{H}_\ell \in C_{\delta,i}}|\widehat{H}_\ell - \breve{H}_i |^2 \Bigg)^{1/2}
 +  \Bigg( \sum_{\ell: \hspace{0.5mm}\widehat{H}_\ell \in C_{\delta,i}}|\breve{H}_i - H_{(\ell)}  |^2 \Bigg)^{1/2}
 $$
 $$
 + \Bigg( \sum_{\ell: \hspace{0.5mm}\widehat{H}_\ell \in C_{\delta,i}} |  H_{(\ell)}- \textnormal{mean}(C_{\delta,i})|^2 \Bigg)^{1/2}
$$
$$
\leq |C_{\delta,i}|^{1/2} \cdot \delta + |C_{\delta,i}|^{1/2} \cdot |\breve{H}_i -\textnormal{mean}(C_{\delta,i}) | = |C_{\delta,i}|^{1/2} \big( \delta + o_{\bbP}(1)\big). 
$$
Consequently, in view of \eqref{e:ICSD} and \eqref{e:ICSD(C_varepsilon)=ICSD(C_delta)}, \eqref{e:ICSD(Cdelta)<small} holds with probability tending towards $1$. This establishes ($i$).

We now show ($ii$). Under \eqref{e:varepsilonb>Delta_min}, observe that there are at least two modes, $\breve{H}_i \neq \breve{H}_j$, with $|\breve{H}_i - \breve{H}_j|<\varepsilon_b$. For the sake of clarity, we focus on the case where $\breve{H}_1$ and $\breve{H}_2$ are the \textit{only} two such values, i.e.,
\begin{equation}\label{e:Hbreve_2-Hbreve_1<varepsilon_b}
 0 < \breve{H}_2 - \breve{H}_1 < \varepsilon_b < \max_{i=2,\hdots,r} \big\{ \breve{H}_{i+1} -  \breve{H}_{i} \big\}. 
 \end{equation} 
 The general case can be worked out based on a tedious adaptation of the argument for the case \eqref{e:Hbreve_2-Hbreve_1<varepsilon_b}.
 
So, under condition \eqref{e:Hbreve_2-Hbreve_1<varepsilon_b}, natural adaptations of the proofs allow us to establish analogous claims to those in Lemmas \ref{p:log_eigenvalues_count} and \ref{p:ball_average_and_proportion} as well as those in the proof of Proposition \ref{t:fixed_epsilon_consistency}. In more detail, analogously to relations \eqref{e:card_rescaled_log-eig=card_H=Hbreve} and \eqref{e:intersection_converges_to_zero}, with probability going to 1 as $n \rightarrow \infty$,
\begin{equation}\label{e:card_rescaled_log-eig=card_H=Hbreve_except_Hbreve1,Hbreve2}
|\widehat{\mathsf{H}}_{p} \cap {\mathcal B}_{\varepsilon_b}(\breve{H}_i)|= \sum_{H\in\mathsf{H}_p}\mathds{1}_{\{H =\breve{H}_i\}}, \quad i= 3,\hdots,r,
\end{equation}
and
\begin{equation}\label{e:card_rescaled_log-eig=card_H=Hbreve1,Hbreve2}
|\widehat{\mathsf{H}}_{p} \cap {\mathcal B}_{\varepsilon_b}(\breve{H}_1)|= |\widehat{\mathsf{H}}_{p} \cap {\mathcal B}_{\varepsilon_b}(\breve{H}_2)| = \sum_{H\in\mathsf{H}_p}\mathds{1}_{\{H =\breve{H}_1 \textnormal{ or }\breve{H}_2\}}.
\end{equation}
 In addition, with the exception of $i = 1$ and $j =2$, for each pair $1\leq i < j\leq r$,
\begin{equation}\label{e:intersection_converges_to_zero_except_Hbreve1,Hbreve2}
\big|\widehat{\mathsf{H}}_{p} \cap {\mathcal B}_{\varepsilon_b}(\breve{H}_i)\bigcap\widehat{\mathsf{H}}_{p} \cap {\mathcal B}_{\varepsilon_b}(\breve{H}_j)\big| = 0,
\end{equation}
again with probability going to 1. In light of the definition \eqref{e:ball_converges_to_proportion}, let
\begin{equation}\label{e:ball_converges_to_proportion_p0}
\widehat{p}_0 = \widehat{p}_0(n):= |\widehat{\mathsf{H}}_p \cap \mathcal{B}_{\varepsilon_b}(\breve{H}_1) \cap \mathcal{B}_{\varepsilon_b}(\breve{H}_2)  | \in \bbN \cup \{0\}, \quad i=1,\hdots,r.
\end{equation}
Then, we conclude that, as $n \rightarrow \infty$,
\begin{equation}\label{e:proportion_ball_converges_to_probability_except_Hbreve1,Hbreve2}
    \frac{\widehat{p}_i(n) }{p(n)} \stackrel{\bbP}\to \pi(\breve{H}_i), \quad i = 3,\hdots,r, \quad \quad \frac{\widehat{p}_0(n) }{p(n)} \stackrel{\bbP}\to \pi(\breve{H}_1) + \pi(\breve{H}_2).
\end{equation}
We further conclude that, as $n \rightarrow \infty$,
\begin{equation}\label{e:mean_of_ball_converges_except_Hbreve1,Hbreve2}
\textnormal{mean} (\widehat{\mathsf{H}}_{p} \cap {\mathcal B}_{\varepsilon}(\breve{H}_i)) \stackrel{\bbP}\to \breve{H}_i, \quad i = 3,\hdots,r,
\end{equation}
and
\begin{equation}\label{e:mean_of_ball_converges_except_Hbreve1,Hbreve2}
\textnormal{mean} \big(\widehat{\mathsf{H}}_{p} \cap {\mathcal B}_{\varepsilon_b}(\breve{H}_1)\big) \stackrel{\bbP}\to  \frac{\pi_1  \breve{H}_1 + \pi_2  \breve{H}_2}{\pi_1 + \pi_2}
 \stackrel{\bbP}\leftarrow \textnormal{mean} \big(\widehat{\mathsf{H}}_{p} \cap {\mathcal B}_{\varepsilon_b}(\breve{H}_2)\big) .
\end{equation}
We now show that, with probability going to 1 as $n \rightarrow \infty$,
 \begin{equation}\label{e:r_hat_estimate_is_consistent_except_Hbreve1,Hbreve2}
        \widehat{r}_{\varepsilon_b}= r-1.
    \end{equation}
In fact, consider the graph induced by an $\varepsilon_b$-threshold $G_{\varepsilon_b}(\widehat{\mathsf{H}}_p)$. By \eqref{e:card_rescaled_log-eig=card_H=Hbreve_except_Hbreve1,Hbreve2} and \eqref{e:card_rescaled_log-eig=card_H=Hbreve1,Hbreve2},  with probability tending to $1$ we can write the decomposition
    \begin{equation}\label{e:G_decomp_disjoint_complete_graphs_except_Hbreve1,Hbreve2}
    G_{\varepsilon_b}(\widehat{\mathsf{H}}_p) = G_0 \cupdot \bigcupdot^{r}_{i=3} G_i
    \end{equation}
(cf.\ \eqref{e:G_decomp_disjoint_complete_graphs}). In \eqref{e:G_decomp_disjoint_complete_graphs_except_Hbreve1,Hbreve2}, $G_i = (E_i,V_i)$, $i = 0,3,\hdots,r$, are complete and pairwise disjoint (sub)graphs.  For each $i$, we define that the vertices in $G_i$ correspond to the points in $ \widehat{\mathsf{H}}_p\cap \mathcal{B}_\varepsilon(\breve{H}_i)$, i.e., we write 
\begin{equation}\label{e:graph_decomp_indexing}
\widehat{H}_\ell \in \widehat{\mathsf{H}}_p\cap \mathcal{B}_\varepsilon(\breve{H}_i) \hspace{1mm} \Leftrightarrow \hspace{1mm} \ell \in V_i.
 \end{equation}
As a consequence of \eqref{e:G_decomp_disjoint_complete_graphs_except_Hbreve1,Hbreve2} and Lemma \ref{l:null_space_of_graph_laplacian}, the non-zero eigenvalues of ${\mathbf L}_\varepsilon(\widehat{\mathsf{H}}_p)$ take values 
\begin{equation}\label{e:p-hat_ell_values_p0}
\widehat{p}_0,\widehat{p}_3\hdots,\widehat{p}_r
\end{equation}
(see \eqref{e:ball_converges_to_proportion} and \eqref{e:ball_converges_to_proportion_p0}), each with multiplicity 
\begin{equation}\label{e:p-hat_ell-1_multiplicity_p0}
\widehat{p}_0-1,\widehat{p}_3-1,\hdots,\widehat{p}_r-1,
\end{equation}
respectively (cf.\ relations \eqref{e:p-hat_ell_values} and \eqref{e:p-hat_ell-1_multiplicity}). So, let $\widehat{p}_{(1)} \leq \widehat{p}_{(2)} \leq \hdots \leq \widehat{p}_{(r-1)}$ be the associated ordered $(r-1)$-tuple. We obtain
\begin{equation}\label{e:eigen_gap_converges_except_Hbreve1,Hbreve2}
\bbP\big(\theta_{r} -\theta_{r-1}  = \widehat{p}_{(1)}\big)\to 1, \quad n \rightarrow \infty.
\end{equation}
On the other hand, by \eqref{e:proportion_ball_converges_to_probability_except_Hbreve1,Hbreve2} and assumption ($A1$) (see expression \eqref{e:difference_of_probabilites_bounded}),
\begin{equation}\label{e:proportion_gap_except_Hbreve1,Hbreve2}
\bbP\Big( \widehat{p}_{(1)} \geq \max_{i = 1,\hdots,r-2}\{\widehat{p}_{(i+1)} -\widehat{p}_{(i)}\}\Big) \to 1,\quad n\to\infty.
\end{equation}
Therefore, from expressions \eqref{e:eigen_gap_converges_except_Hbreve1,Hbreve2} and \eqref{e:proportion_gap_except_Hbreve1,Hbreve2},
\begin{equation}\label{e:P(theta_r+1-theta_r>= max)_p0}
\bbP\Big( \theta_{r} -\theta_{r-1} \geq \max_{i = 1,\hdots, r-2}\{\widehat{p}_{(i+1)} -\widehat{p}_{(i)}\}\Big) \to 1,\quad n\to\infty
\end{equation}
(cf.\ relation \eqref{e:P(theta_r+1-theta_r>= max)}). As a consequence of \eqref{e:p-hat_ell-1_multiplicity_p0}, \eqref{e:proportion_gap_except_Hbreve1,Hbreve2} and \eqref{e:P(theta_r+1-theta_r>= max)_p0}, $\widehat{r}_\varepsilon := \textnormal{argmax}_{i = 1,\hdots,  p(n)-1} \big\{ \theta_{i}  -  \theta_{i-1} \big\} \stackrel{\bbP}\to r-1$ as $n\to\infty$.  This establishes \eqref{e:r_hat_estimate_is_consistent_except_Hbreve1,Hbreve2}.\\

As in the proof of Proposition \ref{t:fixed_epsilon_consistency}, starting from \eqref{e:G_decomp_disjoint_complete_graphs_except_Hbreve1,Hbreve2} and bearing in mind \eqref{e:r_hat_estimate_is_consistent_except_Hbreve1,Hbreve2}, \textbf{HD$\varepsilon$SE} yields $\widehat{r}_{\varepsilon_b} = r - 1$ clusters $\mathcal{C}_{\varepsilon_b}$. Among these clusters, with probability approaching $1$ there is a (nonempty) cluster $C_0 \in \mathcal{C}_{\varepsilon_b}$ with
\begin{equation}\label{e:double_cluster}
    C_0= \widehat{\mathsf{H}}_p\cap \mathcal{B}_{\varepsilon_b}(\breve{H}_1) =\widehat{\mathsf{H}}_p\cap \mathcal{B}_{\varepsilon_b}(\breve{H}_2) =: C_1 \cupdot C_2.
\end{equation}
In \eqref{e:double_cluster}, based on Lemma \ref{p:log_eigenvalues_count} (see \eqref{e:card_rescaled_log-eig=card_H=Hbreve}), for $\varepsilon < \Delta_{\min}$ we can define the asymptotically disjoint clusters $C_i = \widehat{\mathsf{H}}_p\cap \mathcal{B}_{\varepsilon}(\breve{H}_i)$, $i = 1,2$. Then,
\begin{equation}\label{e:ICSD_lower_bound}
\big\{\textnormal{ICSD}(\mathcal{C}_{\varepsilon_b}) \big\}^2 \geq \frac{1}{|C_0|} \sum_{\ell: \hspace{0.5mm} \widehat{H}_{\ell} \in C_0} |\widehat{H}_{\ell}  - \textnormal{mean}(C_0)|^2
= \frac{1}{|C_0|} \sum^{2}_{i=1}\sum_{\ell: \hspace{0.5mm} \widehat{H}_{\ell} \in C_i} |\widehat{H}_{\ell}  - \textnormal{mean}(C_0)|^2.
\end{equation}
Again by Lemma \ref{p:log_eigenvalues_count}, with increasing probability the wavelet log-eigenvalues $\widehat{H}_{\ell} \in C_0$ coalesce around either $\breve{H}_1$ or $\breve{H}_2$. More specifically, 
for any arbitrarily small $\eta > 0$, as $n \rightarrow \infty$,
\begin{equation}\label{e:max_min_|Hhat_ell-H-breve|}
\max_{\ell: \hspace{0.5mm} \widehat{H}_{\ell} \in C_0} \min_{i=1,2} \big|\widehat{H}_{\ell} - \breve{H}_i \big| < \frac{\eta}{2}
\end{equation}
with increasing probability. Also with increasing probability, \eqref{e:mean_of_ball_converges_except_Hbreve1,Hbreve2} implies that
\begin{equation}\label{e:|mean(C0)-pi1H1+pi2H2/pi1+pi2|_bound}
\Big| \textnormal{mean}(C_0) - \Big( \frac{\pi_1 \breve{H}_1 + \pi_2 \breve{H}_2}{\pi_1 + \pi_2} \Big)\Big| < \frac{\eta}{2}.
\end{equation}
By  relations \eqref{e:max_min_|Hhat_ell-H-breve|} and \eqref{e:|mean(C0)-pi1H1+pi2H2/pi1+pi2|_bound}, as well as the triangle inequality, 
$$
\big|\widehat{H}_{\ell}  - \textnormal{mean}(C_0) \big| \geq \Bigg| \Big| \breve{H}_i - \Big( \frac{\pi_1 \breve{H}_1 + \pi_2 \breve{H}_2}{\pi_1 + \pi_2}\Big)  \Big| 
- \Big| (\widehat{H}_\ell - \breve{H}_i) + \Big( \frac{\pi_1 \breve{H}_1 + \pi_2 \breve{H}_2}{\pi_1 + \pi_2}\Big) - \textnormal{mean}(C_0)\Big| \Bigg| 
$$
\begin{equation}\label{e:|H-hat_ell-mean(C_0)|}
\geq \widetilde{C} \cdot \Big| \breve{H}_i - \Big( \frac{\pi_1 \breve{H}_1 + \pi_2 \breve{H}_2}{\pi_1 + \pi_2}\Big)  \Big|,
\end{equation}
where the last inequality holds for some fixed constant $\widetilde{C} > 0$ for any small enough $\eta$. In addition, relation \eqref{e:proportion_ball_converges_to_probability_except_Hbreve1,Hbreve2} implies that, as $n \rightarrow \infty$, 
\begin{equation}\label{e:|Ci|/|C0|_conv}
\frac{\widehat{p}_i}{\widehat{p}_0} = \frac{|C_i|}{|C_0|} \stackrel{\bbP}\rightarrow \frac{\pi_i}{\pi_1 + \pi_2}, \quad i = 1,2. 
\end{equation}
By relations \eqref{e:ICSD_lower_bound}, \eqref{e:|H-hat_ell-mean(C_0)|} and \eqref{e:|Ci|/|C0|_conv}, we conclude that, with probability going to 1 as $n \rightarrow \infty$, 
$$
\big\{\textnormal{ICSD}(\mathcal{C}_{\varepsilon_b}) \big\}^2 \geq   \widetilde{C} \cdot \sum^{2}_{i=1} \frac{\pi_i}{\pi_1 + \pi_2}  \cdot \Big|\breve{H}_i  - \frac{\pi_1  \breve{H}_1 + \pi_2 \breve{H}_2}{\pi_1 + \pi_2} \Big|^2.
$$
Hence, \eqref{e:bad_ICSD_not_to_zero} holds. This establishes ($ii$), completing the proof. $\Box$\\
\end{proof}

\begin{remark}\label{r:varepsilon=Delta}
Note that the choice $\varepsilon = \Delta_{\min}$ is excluded from the statement of Lemma \ref{p:ICDS_to_zero}. This is so because relation \eqref{e:log_eigenvalue_bound1} does not allow us to establish whether or not the clusters $\widehat{H}_p \cap {\mathcal B}_{\varepsilon}(\breve{H}_i)$, $i = 1,\hdots, r$, provide an asymptotic partition of the set $\widehat{H}_p$. Making such a claim would require a more refined understanding of the fluctuations of wavelet log-eigenvalues than the one provided in the proof of Theorem \ref{t:main_theorem_discrete}.
\end{remark}

%\medskip

We are now in a position to prove the main mathematical result of the paper, i.e., Theorem \ref{t:WRMSM_consistency}.\\

\noindent{\sc Proof of Theorem \ref{t:WRMSM_consistency}}:  For each $\varepsilon \in \mathcal{E}$, let $\mathcal{C}_{\varepsilon}$ be the collection of clusters obtained by means of \textbf{HD$\varepsilon$SE}. Define the subsets $\mathcal{E}_g := \mathcal{E} \cap (0,\Delta_{\min}) \neq \emptyset$ (see \eqref{e:there_is_a_good_epsilon}) and $\mathcal{E}_b:=\mathcal{E} \cap (\Delta_{\min},\infty)$ (using subscripts $g$ and $b$ as in the initials of ``good" and ``bad", respectively). Without loss of generality, assume $\mathcal{E}_b \neq \emptyset$. We claim that
\begin{equation}\label{e:model_selected_is_good}
  \bbP(\varepsilon_{ms}\in \mathcal{E}_g)\to 1,\quad n\to \infty.
\end{equation}
In fact, note that
\begin{equation}\label{e:max(Eg,Eb)<infty}
\max \big\{\textnormal{card}(\mathcal{E}_g),\textnormal{card}(\mathcal{E}_b )\big\} \leq \textnormal{card}(\mathcal{E} ) < \infty.
\end{equation}
By Lemma \ref{p:ICDS_to_zero} and \eqref{e:max(Eg,Eb)<infty}, as $n \rightarrow \infty$,
\begin{equation}\label{e:maxICSD(Eg)->0,minICSD(Eb)_bounded_below}
\max_{\varepsilon_g \in {\mathcal E}_g} \textnormal{ICSD}(\mathcal{C}_{\varepsilon_{g}})\stackrel{\bbP}\to 0, \quad \min_{\varepsilon_b \in {\mathcal E}_b} \textnormal{ICSD}(\mathcal{C}_{\varepsilon_b})> \Upsilon + o_{\bbP}(1)
\end{equation}
for some constant $\Upsilon > 0$. Recall that the (random) threshold $\varepsilon_{ms}$ is chosen such that $\textnormal{ICSD}(\mathcal{C}_{\varepsilon_{ms}}) = \min_{\varepsilon\in\mathcal{E}}\{\textnormal{ICSD}(\mathcal{C}_\varepsilon)\}.$ Then, as a consequence of \eqref{e:maxICSD(Eg)->0,minICSD(Eb)_bounded_below}, relation \eqref{e:model_selected_is_good} holds.

We now turn to \eqref{e:good_epsilon_consistency_ms}. In view of \eqref{e:model_selected_is_good}, we can further write 
\begin{equation}\label{e:varepsilon_ms=sum}
\varepsilon_{ms} = \sum_{\varepsilon_{g} \in {\mathcal E}_g} \varepsilon_{g} \hspace{1mm} \indicator_{\{ \varepsilon_{g} \hspace{0.5mm} = \hspace{0.5mm}\textnormal{argmin}_{\varepsilon \in {\mathcal E}} \textnormal{ICSD}(\mathcal{C}_\varepsilon)\}}
\end{equation}
with increasing probability as $n \rightarrow \infty$. Also, as a consequence of Proposition \ref{t:fixed_epsilon_consistency} (see \eqref{e:good_epsilon_consistency}) and \eqref{e:max(Eg,Eb)<infty}, for any $\eta >0$, as $n\to\infty$,
    \begin{equation}\label{e:good_epsilon_consistency_Eg}
    \bbP\Bigg(  \bigcap_{\varepsilon_g \in {\mathcal E}_g} \Big(\{\widehat{r}_{\varepsilon_g} = r\} \bigcap \Big\{  \max_{i = 1,\hdots, r}| \widehat{\breve{H}}_i -\breve{H}_i| < \eta \Big\}  \bigcap \Big\{ \max_{i = 1,\hdots, r}| \widehat{\pi}_i -\pi(\breve{H}_i)| < \eta \Big\} \Big) \Bigg) \to 1.
    \end{equation}
Relation \eqref{e:good_epsilon_consistency_ms} is now a consequence of \eqref{e:varepsilon_ms=sum} and \eqref{e:good_epsilon_consistency_Eg}, which completes the proof. $\Box$\\

\section{More on graph Laplacians}

In this section, we recap the main lemma on spectral graph theory. The lemma establishes the basic connection between the spectral information of the graph Laplacian and the topology of the graph. In particular, it provides the mathematical foundation for spectral clustering.  The proof of the lemma can be found, for instance, in Orejola \cite{orejola:2024}. In the context of this paper, the lemma is used in the proof of Proposition \ref{t:fixed_epsilon_consistency}.

\begin{lemma}\label{l:null_space_of_graph_laplacian}
Let $G$ be the disjoint union of complete graphs $G_i =(E_i,V_i)$, $i = 1,\hdots,r$, and let $p_i$ be the number of vertices in the graph $G_i$. Then, ${\mathbf L}(G)$ has a zero eigenvalue with multiplicity $r$, whereas the non-zero eigenvalues of ${\mathbf L}(G)$ take values $p_1,\hdots,p_r$ each with multiplicity $p_1-1,\hdots,p_r-1$ respectively. Additionally, let $\mathbf{u}_i$ be the eigenvectors corresponding to the zero eigenvalue of ${\mathbf L}(G)$, and let
\begin{equation}\label{e:spectral_rep}
   {\mathbf U} = \big(    \mathbf{u}_1,\hdots , \mathbf{u}_r \big).
\end{equation}
For each $\ell=1,\hdots,p$, let $\mathbf{y}_\ell$ denote the $\ell$--th row of ${\mathbf U}$. Then, for each $i=1,\hdots,r$ and any $\ell,\ell' \in V_i$, $\mathbf{y}_\ell$ = $\mathbf{y}_{\ell'}$. Moreover, if $\ell' \in V_i$ and $\mathbf{y}_{\ell'} \neq \mathbf{y}_\ell$ then $\ell \notin V_i$.
\end{lemma}

%\begin{proof} As in the proof of Lemma \ref{l:union_of_complete_graphs}, let $G$ be the union of $G_1,\hdots,G_r$. Also, let $p = \sum^{r}_{i=1}p_i$. With $A(G)$ as in \eqref{e:adj_mat}, without loss of generality suppose (up to a relabeling of the vertices), the graph laplacian of $G$ can be expressed as a block matrix consisting sub-graph Laplacians,
%$$
% L(G) = \begin{bmatrix}
% L(G_1) &  & &\\
%&  L(G_2) &  &\\
%  &  & \ddots& \\
%   &  & & L(G_r)\\
%\end{bmatrix}
%$$
%where $L(G_i) = D(G_i)-A(G_i)$. With $L(G)$ block diagonal, the characteristic polynomial of $L(G)$ can be written as $\det(L(G) - \lambda I) = \prod_{i=1}^r\det(L(G_i) - \lambda I_{p_i\times p_i})$. With each $G_i$ simple, by lemma \ref{l:graph_laplacian_complete}, for $i = 1,\hdots,r$, $\det(L(G_i) - \lambda I_{p_i\times p_i})$ equal zero if and only if $\lambda = 0$ or $\lambda = p_i$. Moreover, as in lemma \ref{l:graph_laplacian_complete}, $\mathbf{v}_i^\top L(G_i) \mathbf{v}_i = 0$  if and only if $\mathbf{v}_i$ is a scalar multiple of $[1,\hdots,1]^\top \in \mathbb{R}^{p_i}$.
%Therefore, the non-zero eigenvalues of $L(G)$ take values $p_1,\hdots,p_k$ and $L(G)$ has a zero eigenvalue with multiplicity $k$ with corresponding eigenvectors  ${\mathbf u}_i \in \bbR^n$ with $[{\mathbf u}_i]_j = 1$ for $j \in \{\sum_{\ell = 1}^{i-1}p_i,\hdots,\sum_{\ell = 1}^{i}p_i\} $ and $[{\mathbf u}_i]_j = 0$  otherwise. Consequently, $[{\mathbf u}_i]_j = \mathds{1}_{j \in V_i}$. Therefore, for any $k,l \in V_i$, $\mathbf{y}_k = \mathbf{y}_l$. This completes the proof. $\Box$
%\end{proof}

\section{The empirical spectral distribution of large wavelet random matrices}\label{s:wavelet_e.s.d.}

The following is the fundamental result on the bulk behavior of wavelet log-eigenvalues in a moderately high-dimensional framework.
\begin{theorem}\label{t:main_theorem_discrete}
(Abry, Didier et al.\ \cite{abry:didier:orejola:wendt:2024}, Theorem 3.3) Suppose assumptions ($A1-A5$) and ($W1-W3$) hold. For  $\widehat{H}_1,\hdots,\widehat{H}_p$ as in \eqref{e:scaling_estimate}, let
\begin{equation}\label{e:ecdf_discrete}
F_{ {\boldsymbol {\mathcal W}}_n } (\upsilon) := \frac{1}{p(n)}\sum_{\ell=1}^{p(n)}\mathds{1}_{  \{ 2 \cdot \widehat{H}_\ell + 1 \leq \upsilon\}}, \quad \upsilon \in \bbR.
\end{equation}
Then, the e.s.d.\ $F_{ {\boldsymbol {\mathcal W}}_n }$ converges weakly, in probability, to $F_{2\mathcal{H}+1}$. In other words, for any $g \in C_{b}(\bbR)$ and for any $\varepsilon > 0$,
\begin{equation}\label{e:main_theorem_discrete}
\lim_{n \rightarrow \infty}\bbP\Big(\Big|\int_{\bbR} g(\upsilon) F_{ {\boldsymbol {\mathcal W}}_n }(d\upsilon) -\int_{\bbR} g(\upsilon) F_{2\mathcal{H}+1}(d\upsilon)\Big| > \varepsilon \Big)= 0.
\end{equation}
\end{theorem}

\section{The $k$-means algorithm}\label{s:k-means}

In this section, for the readers' convenience we recap the $k$-means algorithm as pseudocode, and state its relevant properties. For more information about $k$-means, we direct readers to Hastie et al.\ \cite{hastie:tibshirani:friedman:2009}.

%\begin{lstlisting}[frame=single,escapechar=\%]
%$\boldsymbol{k}\mathbf{-means}$:
%$\mathbf{Input}$: $\kappa \in\mathbb{N}$,  $\mathcal{X} = \{ \x_1,..., \x_p\}$, where $\x_i \in \mathbb{R}^d$, $i=1,\hdots,p$.
%$\bullet$ Select any collection of distinct points $\{c_1,\hdots,c_{\kappa}\}\subseteq \mathcal{X}$.
%$\ \mathbf{1)}$ For each $\x \in\mathcal{X}$ compute $\textnormal{dist}_x := \{ \|\x - c_1\|,\hdots,\| \x-c_{\kappa}\|\}$.
%$\ \mathbf{2)}$ Compute clusters $C_i = \{ \x \in \mathcal{X}\ : \ \|\x - c_i\| = \min( \textnormal{dist}_{\x})\}$, for $i=1,\hdots,\kappa$.
%$\ \mathbf{3)}$ For $i=1,\hdots,\kappa$, let $c_i = \textnormal{mean}(\mathcal{C}_i)$.
%$\bullet$ Repeat $\mathbf{1}$, $\mathbf{2}$ and $\mathbf{3}$ until $c_i$ does not change, for each $i=1,\hdots, \kappa$.
%$\mathbf{Output}$: Clusters $C_1,\hdots,C_{\kappa}$.
%\end{lstlisting}
%

%%%%%%%%%%%%%%%%%%%%%%%%%%%%%%%%%%%%%%%%%%%%%%%%%%%%%%%%%%%%%%%%%%%%%%%%%%
%\vspace{-0.5cm}
{\small
\begin{center}
\begin{tabular}{|l|}
\hline \\
\multicolumn{1}{|c|}{pseudocode for the $k$-\textbf{means} algorithm}\\ 
\\
\hline\\
\textbf{Input}: $\kappa \in\mathbb{N}$,  $\mathcal{X} = \{ \x_1,..., \x_p\}$, where $\x_i \in \mathbb{R}^d$, $i=1,\hdots,p$.\\
\\
$\bullet$ Select any collection of distinct points $\{c_1,\hdots,c_{\kappa}\}\subseteq \mathcal{X}$.\\
\\
\textbf{Step 1}: For each $\x \in\mathcal{X}$ compute $\textnormal{dist}_x := \{ \|\x - c_1\|,\hdots,\| \x-c_{\kappa}\|\}$. \\
\\
\textbf{Step 2}: Compute clusters $C_i = \{ \x \in \mathcal{X}\ : \ \|\x - c_i\| = \min( \textnormal{dist}_{\x})\}$, for $i=1,\hdots,\kappa$. \\
\\
\textbf{Step 3}: For $i=1,\hdots,\kappa$, let $c_i = \textnormal{mean}(\mathcal{C}_i)$.\\
\\
$\bullet$ Repeat $\mathbf{1}$, $\mathbf{2}$ and $\mathbf{3}$ until $c_i$ does not change, for each $i=1,\hdots, \kappa$.\\
\\
\textbf{Output}:  Clusters $C_1,\hdots,C_{\kappa}$.\\ 
\\ \hline
\end{tabular}
\end{center}
}

\vspace{2mm}

%\GDcomment{Is the stopping criterion included in the description of $k$-means provided in Hastie et al?} \OOcomment{Yes, see page 510.}

The $k$-means algorithm inherently produces convex clusters due to the nature of its objective function and optimization process. Note that, in practice, the convergence criterion for iterative procedure is often presupposed by a maximum iteration condition. That is, the algorithm terminates if the clusters means, $c_i$, do not change or a maximum number of iterations is meet. However, for our theoretical consideration a maximum iteration condition is not necessary.

The next lemma establishes the simple fact that if one performs $k$-means, setting $\kappa=r$, on a collection of points that can take of only $r$ distinct values, then one obtains clusters each associated to one of the $\kappa$-distinct points.

\begin{lemma}\label{l:k-means_converges}
Fix $p \in \bbN$ and let $r < p$. Suppose that $\mathcal{X}$ is a collection of $p$ points in $\mathbb{R}^d$ taking only $r$ distinct (vector) values 
\begin{equation}\label{e:x'1,...,x'r}
\{{\mathbf x}'_1,\hdots,{\mathbf x}'_r\}. 
\end{equation}
Then, setting $\kappa =r$, an application of the $k$-means algorithm yields the clusters
\begin{equation}\label{e:C1,...,Cr}
C_i = \big\{ {\mathbf x} \in \mathcal{X} : {\mathbf x} = {\mathbf x}'_i \big\}, \quad i = 1,\hdots,r.
\end{equation}
\end{lemma}
\begin{proof}
Recall that the $k$-means algorithm is initialized based on a collection of distinct points $\{{\mathbf c}_1,\hdots,{\mathbf c}_{r}\} \subseteq \mathcal{X}$. In view of \eqref{e:x'1,...,x'r}, up to relabeling the indices we can assume that 
\begin{equation}\label{e:ci=x'i}
{\mathbf c}_i ={\mathbf x}'_i, \quad i=1,\hdots,r. 
\end{equation}
Moreover, again in view of \eqref{e:x'1,...,x'r}, it follows that, for every $i = 1,\hdots,r$,  $\|{\mathbf x}-{\mathbf c}_i\| =0$ if and only if ${\mathbf x} ={\mathbf x}'_i$. Therefore, for each $i=1,\hdots,r$, $ C_i = \{{\mathbf x} \in \mathcal{X} : {\mathbf x}  = {\mathbf x}'_i\}$. Consequently, following the $k$-means algorithm, the new cluster means are given by 
\begin{equation}\label{e:x'i=ci}
\textnormal{mean}(C_i) = {\mathbf x}'_i = {\mathbf c}_i, \quad i=1,\hdots,r. 
\end{equation}
As a consequence of \eqref{e:ci=x'i} and \eqref{e:x'i=ci}, the cluster means did not change. Hence, the $k$-means algorithm stops, yielding the clusters \eqref{e:C1,...,Cr}. This completes the proof. $\Box$\\
\end{proof}

%\begin{lemma}\label{l:k-means_converges}
 %Let $G$ be the disjoint union of simple graphs $G_1 =(E_1,V_1),\hdots,G_r = (E_r,V_r)$. Also, for $i = 1,\hdots,r$, let $p_i$ be the number of vertices in the graph $G_i$, and let $p=p_1+\hdots+p_r$. Let $U$ be as in \eqref{e:spectral_rep}, and let $\mathbf{y}_j$ denote the $j$th row of $U$. Then, an application of the $k$-means algorithm, setting $k=r$, to the set $\{\mathbf{y}_1,\hdots,\mathbf{y}_p\}$ yields clusters
 %$$
 %C_1 =\{\mathbf{y}_\ell \ : \ \ell \in V_1 \},\hdots ,C_\ell =\{\mathbf{y}_i \ : \ \ell \in V_r \}.
 %$$
%\end{lemma}

%\begin{proof}
% Consider any collection of distinct points $\{c_1,\hdots,c_r\} \subset \{\mathbf{y}_1,\hdots,\mathbf{y}_p\}$ as in the initialization of the $k$-means algorithm. Observe that from Lemma \ref{l:null_space_of_graph_laplacian}, for each $i=1,\hdots,r$, if $k,\ell\in V_i$ then $\mathbf{y}_i = \mathbf{y}_\ell$. Additionally, if $k\in V_i$ and $\mathbf{y}_k \neq \mathbf{y}_\ell$ then $\ell \notin V_i$. Therefore, for each $i=1,\hdots,r$, $\mathbf{y}_k \in C_i$ if and only if $\mathbf{y}_k = c_i$
%\end{proof}

\bibliography{highdim}

\end{document}